\documentclass[fleqn,usenatbib]{mnras}
\usepackage{newtxtext,newtxmath}
\usepackage{hyperref} 
\usepackage[T1]{fontenc}
\usepackage{ae,aecompl}
\usepackage{graphicx,epsfig}	
\usepackage{amsmath}           	
\usepackage{booktabs} 
\usepackage[dvipsnames]{xcolor}
\usepackage{url}

\usepackage{stfloats} 

\newcommand{\ppxf}{{\tt pPXF}\,}
\newcommand{\msun}{M$\mathrm{_\odot}$}
\newcommand{\mhalo}{$M\mathrm{_{halo}}$}

\newcommand{\Mstar}{$M_{*}$\,}

\newcommand{\afe}{[$\alpha$/Fe]}

\newcommand{\kms}{km~s$^{-1}$}

\newcommand{\re}{$R_{\rm e}$}

\title[SFHs of quiescent UDGs, environ and GC richness]{The star formation histories of quiescent ultra-diffuse galaxies and their dependence on environment and globular cluster richness}
\author[A. Ferr\'e-Mateu et  al.]{Anna Ferr\'e-Mateu$^{1,2,3}$\thanks{E-mail: aferre@iac.es (AFM)}, Jonah S. Gannon$^{3,4}$, Duncan A. Forbes$^{3,4}$, Maria Luisa Buzzo$^{3,4}$, 
\newauthor Aaron J. Romanowsky$^{5,6}$ and Jean P. Brodie$^{3,4}$\\
$^{1}$ Instituto Astrofisica de Canarias, Av. Via Lactea s/n, E38205 La Laguna, Spain\\
$^{2}$ Departamento de Astrofisica, Universidad de La Laguna, E-38200, La Laguna, Tenerife, Spain\\
$^{3}$ Centre for Astrophysics \& Supercomputing, Swinburne University of Technology, Hawthorn VIC 3122, Australia\\
$^{4}$ARC Centre of Excellence for All Sky Astrophysics in 3 Dimensions (ASTRO 3D), Australia\\
$^{5}$ Department of Physics \& Astronomy, San José State University, One Washington Square, San Jose, CA 95192, USA\\
$^{6}$ Department of Astronomy \& Astrophysics, University of California Santa Cruz, 1156 High Street, Santa Cruz, CA 95064, USA
\\
}

\date{Awaiting referee comments after minor revision; Received April 2023}
\pubyear{2023}

\begin{document}
\label{firstpage}
\pagerange{\pageref{firstpage}--\pageref{lastpage}}
\maketitle

\begin{abstract}
We derive the stellar population parameters of 11 quiescent ultra-diffuse galaxies (UDGs) from Keck/KCWI data. We supplement these with 14 literature UDGs, creating the largest spectroscopic sample of UDGs to date (25). We find a strong relationship between their $\alpha$-enhancement and their star formation histories: UDGs that formed on very short timescales have elevated [Mg/Fe] abundance ratios, whereas those forming over extended periods present lower values. Those forming earlier and faster are overall found in high-density environments, being mostly early infalls into the cluster. No other strong trends are found with infall times. We analyze the stellar mass--metallicity, age--metallicity and [Mg/Fe]--metallicity relations of the UDGs, comparing them to other types of low mass galaxies. Overall, UDGs scatter around the established stellar mass--metallicity relations of classical dwarfs. We find that GC-rich UDGs have intermediate-to-old ages, but previously reported trends of galaxy metallicity and GC richness are not reproduced with this spectroscopic sample due to the existence of GC-rich UDGs with elevated metallicities. In addition, we also find that a small fraction of UDGs could be ‘failed-galaxies’, supported by their GC richness, high alpha-abundance, fast formation timescales and that they follow the mass--metallicity relation of $z\sim$2 galaxies. Finally, we also compare our observations to simulated UDGs. We caution that there is not a single simulation that can produce the diverse UDG properties simultaneously, in particular the low metallicity failed-galaxy like UDGs.
\end{abstract}


\section{Introduction}\label{section:intro}
Ultra-diffuse galaxies (UDGs) are a type of low surface brightness (LSB; \citealt{Sandage1984}) galaxy, broadly defined by being extremely faint ($\mu_{0,g}$>24 mag\,arcsec$^{-2}$), presenting a relatively large half-light radius (\re > 1.5 kpc) and exhibiting dwarf-like stellar masses (log(\Mstar/\msun)$\sim$7--8.5; see \citealt{vanNest2022} for an in-depth discussion on UDG definition and \citealt{Watkins2023} for limitations on the use of this broad definition). While galaxies fitting this definition had been previously discovered (e.g., \citealt{Impey1988}), limitations in the telescope facilities and data techniques at the time prevented a further characterization of their nature. 

Nowadays, with instrumentation and techniques specifically developed to push the limits reachable in LSB studies, UDGs were re-discovered in 2015 by \citet{vanDokkum2015a} and \citet{Koda2015}. They were found by the hundreds in the nearby Coma cluster and since then they have been found to exist in all environments. They are seen in other galaxy clusters besides Coma (e.g., \citealt{Mihos2015}; \citealt{Mancera-Pina2018}; \citealt{Lim2020};  \citealt{Iodice2020}), in groups (e.g., \citealt{Trujillo2017}; \citealt{Muller2018}; \citealt{Forbes2019}; \citealt{Forbes2020b}; \citealt{Mareleu2021}), and in the field (\citealt{MartinezDelgado2016}; \citealt{Bellazzini2017}; \citealt{Leisman2017}; \citealt{Greco2018}; \citealt{Roman2019}; \citealt{Barbosa2020}; \citealt{Zaritsky2022}). A transition in the overall properties of cluster $vs.$ field UDGs is seen, with UDGs in clusters being mostly red, spheroidal and quenched, while those in the field tend to be bluer, more irregular and gas-rich (e.g., \citealt{vanderBurg2016}; \citealt{Leisman2017}; \citealt{Roman2017a}). Moreover, they have been found to exist at least up to redshift $z\sim$0.8 (e.g., \citealt{Janssens2017}; \citealt{Janssens2019}; \citealt{Lee2020}; \citealt{Carleton2023}; \citealt{Ikeda2023}). 

In order to better understand how UDGs formed, a large effort has been undertaken in the past few years to characterize the main properties of these LSB galaxies. Their dark matter content has been one of the most revealing properties. Some UDGs have inferred dark matter halos similar to those found in classical dwarf galaxies at the same stellar mass (log(\mhalo/\msun) $\lesssim$10.5; \citealt{BeasleyT2016}; \citealt{Amorisco2018a}; \citealt{Gannon2021}; \citealt{Iodice2020}), while other UDGs show evidence of more massive dark matter halos (log(\mhalo/\msun) $\gtrsim$10.5; \citealt{Beasley2016}; \citealt{Toloba2018}; \citealt{vanDokkum2019}; \citealt{Forbes2021}; \citealt{Gannon2022}). In some rare cases, they have been found to host no dark matter at all, a very disputed result that still attracts a lot of attention (e.g., \citealt{vanDokkum2018}; \citealt{Wasserman2018};  \citealt{Trujillo2019}; \citealt{vanDokkum2019}; \citealt{Danielli2020}; \citealt{Montes2020}; \citealt{Montes2021}; \citealt{Shen2021}). 

Some UDGs have been found to host rich systems of globular clusters (GCs; e.g., \citealt{Beasley2016}; \citealt{vanDokkum2017}; \citealt{Amorisco2018b}; \citealt{Lim2018}; \citealt{vanDokkum2018}; \citealt{Forbes2019}; \citealt{Forbes2020c}; \citealt{Mueller2021}; \citealt{Saifollahi2021}), whereas other UDGs can be extremely GC-poor (e.g., \citealt{Forbes2020c}; \citealt{Gannon2022b}; \citealt{Jones2022}; \citealt{Saifollahi2022}). There is a tight correlation between the dark matter halo mass and the total number of GCs in normal galaxies (e.g., \citealt{Spitler2009,Harris2013,Burkert2020}), which UDGs also seem to follow (e.g., \citealt{Gannon2022}; \citealt{Zaritsky2023}). GC-poor UDGs are thus likely to reside in dark matter halos similar to those found for classical dwarf galaxies, while GC-rich ones may reside in more massive dark matter halos (e.g. \citealt{Sifon2018}; \citealt{Gannon2022b}). 

Evidence of the latter is seen in the kinematic studies of several GC-rich UDGs. Measuring their dynamical masses, they were found to be consistent with halo-profiles for massive dark matter halos log(\mhalo/\msun)$\gtrsim$10.5 (e.g., \citealt{Beasley2016}; \citealt{vanDokkum2016}; \citealt{Toloba2018}; \citealt{Forbes2021}, \citealt{Gannon2020G}; \citealt{Gannon2022}). However, much is yet to be unveiled for the GC-poor UDGs and their dark matter halos. Altogether, from these clearly distinctive properties, two main different types of UDGs have been proposed to co-exist. Several mechanisms have been suggested to create each class of UDG, with different sets of cosmological simulations trying to reproduce them.

Primarily, many propose that UDGs are classical dwarf galaxies that are `puffed up' by some mechanism into the large size observed, which we will refer to as the \textit{puffy dwarf formation scenario}. This can be attributed to either internal feedback (e.g., \citealt{DiCintio2017}; \citealt{Chan2017}), high halo spins (e.g., \citealt{Amorisco2016}; \citealt{Rong2017}; \citealt{Liao2019}), or external effects such as quenching, tidal forces, or ram pressure stripping (e.g., \citealt{Safarzadeh2017}; \citealt{Ogiya2018}; \citealt{Carleton2019}; \citealt{Sales2020}; \citealt{Tremmel2020}; \citealt{Benavides2021}; \citealt{Jones2021}; \citealt{Junais2022}). Combinations of these effects may also occur, although they do not need to happen simultaneously (e.g., \citealt{Jiang2019}; \citealt{Martin2019}; \citealt{Liao2019}; \citealt{Trujillo-Gomez2022}; \citealt{Watkins2023}). 

Another school of thought describes UDGs as a remnant dark matter halo from the early Universe. An ancient, dark matter-dominated, proto-galaxy could have been quenched very early on, preventing its subsequent expected evolution into a regular galaxy. This prevented the galaxy from acquiring the stellar mass expected given its original dark matter halo, explaining the massive dark matter halo it still resides in (e.g., \citealt{Yozin2015}; \citealt{Peng2016}; \citealt{Rong2017}; \citealt{Danieli2022}). This scenario was originally named `failed Milky Way-galaxies' \citep{vanDokkum2015a}, although it has since been demonstrated that these galaxies can not reside in dark matter halos quite as massive as the Milky Way (see e.g., \citealp{Sifon2018}). Hence, we will use the simplified term \textit{failed-galaxy scenario} hereafter. These UDGs, while still being LSB dwarfs in terms of their stellar masses, are expected to reside in massive dark matter halos (e.g., \citealt{Harris2013}; \citealt{Burkert2020}), thus are expected to be GC-rich systems. However, they have been hardly reproduced by any cosmological simulations yet.
 
One of the best ways to investigate this presumed dichotomy in the origin of UDGs comes from the analysis of their stellar population properties, as precise and distinct predictions for each type are expected. Some works tackled this by using SED fitting to derive some stellar population properties of a large number of UDGs at once (e.g., \citealt{Pandya2017}; \citealt{Barbosa2020}; \citealt{Buzzo2022}). However, SED fitting is not as well constrained as spectroscopy, and only a few spectroscopic attempts have been carried to date for UDGs. This has been done for either limited samples of UDGs (mostly in clusters, e.g., \citealt{Gu2017}; \hypertarget{FM+18}{\citealt{Ferre-Mateu2018a}} (FM+18 hereafter); \citealt{Ruiz-Lara2018}; \citealt{Chilingarian2019}), or for some individual cases, including UDGs in less dense environments (e.g., \citealt{Fensch2019}; \citealt{Martin-Navarro2019}; \citealt{Mueller2021}; \citealt{Villaume2022}). Sometimes they have been stacked to enhance the quality of the spectra, as done for the star-forming isolated UDGs from SDSS \citep{Rong2020a}. 

Under the puffy dwarf-like scenario, UDGs are expected to have mostly extended star formation histories (SFHs) and thus younger ages ($\sim$6--9\,Gyr), with metallicities similar to classical dwarfs ([M/H]$\sim -$0.8\,dex; e.g., \hyperlink{FM+18}{FM+18}; \citealt{Ruiz-Lara2018}; \citealt{Chilingarian2019}; \citealt{Martin-Navarro2019}; \citealt{Rong2020b}). Failed-galaxy UDGs would instead expected to have stellar populations more similar to GCs, with old ages ($\sim$10--12\,Gyr) as they were quenched very early on. In such case, their metallicities are also expected to be much lower than the puffy dwarf-like UDGs ([M/H]$\sim -$1.5\,dex; e.g., \citealt{Naidu2022}).

One intriguing property that has not been tackled much yet for UDGs, and that is not possible to obtain from SED fitting, is the [$\alpha$/Fe] abundance ratio (or [Mg/Fe], depending on the work and how it is measured). From the few estimates available, UDGs seem to present elevated ratios ($\gtrsim$0.3\,dex; e.g., \hyperlink{FM+18}{FM+18}; \citealt{Ruiz-Lara2018}; \citealt{Rong2020a}). In the case of the isolated UDG DGSAT~I, it exhibits the most extreme [Mg/Fe] ever found in any type of galaxy \citep{Martin-Navarro2019}. In a canonical view, the $\alpha$-enhancement has been suggested to be a cosmic-clock, as each element is formed at different timescales (e.g., \citealt{Matteucci1994}; \citealt{Thomas2005}; \citealt{deLaRosa2011}; \citealt{McDermid2015}). Galaxies forming on very short timescales present elevated [Mg/Fe], as the first yields to be produced by core-collapse are those of $\alpha$-elements such as magnesium. As time evolves, galaxies with more extended SFHs decrease their [Mg/Fe] values (typically down to solar abundance ratios) as the inter-stellar medium is polluted by the ejection of iron from Type Ia supernovae. Therefore galaxies with elevated [Mg/Fe] but extended SFHs, as found for many UDGs, is a surprising result under the current galaxy formation paradigm.

The challenge in all these spectroscopic studies is that acquiring good quality spectra for stellar population analysis requires a large time investment with 10m class telescopes. A balance between the quality of the spectra and the number of UDGs observed in the sample must therefore be found. Moreover, most of the studied samples are clearly biased towards cluster environments, leaving field or group UDGs mostly unexplored. To work on this, we have carried several observational campaigns to obtain moderate quality spectra (S/N$\sim$15--20) for a large sample of UDGs in different environments, although with a heavy preference towards Perseus UDGs. As a result, adding these new UDGs (11) to previously published UDGs (14), we nearly double the number of quiescent UDGs with spectroscopic measurements. We consider UDGs to be quiescent when they have no signs of emission lines and do not exhibit ongoing star formation. We note that this does not exclude UDGs having experienced recent star formation (e.g., in the last few Gyr). 

We have thus created the largest spectroscopic sample of quiescent UDGs with stellar population properties. For most of the new UDGs, a dynamical study of their masses, velocity dispersion and dark matter content has been already published (\citealt{Gannon2020G}, \citealt{Gannon2021}; \citealt{Forbes2021}; \citealt{Gannon2022} and \citealt{Gannon2022b}). 

\begin{table*}
\centering
\label{table:bonafide}                      
\begin{tabular}{c |c c c c c c}   
\toprule      
\textbf{Galaxy}  & \textbf{Env.} &  V$_r$  & \textbf{\Mstar}          & \textbf{\re} & \textbf{GC-richness} & \textbf{Data} \\  
                 &               &   \kms  & ($\times 10^{8}$\msun)   &    (kpc)     &   \textbf{classification}                   &  \\
\midrule                             
\midrule                          
This work                      &                          &         &        &         &             & \\
\midrule                                                      
VCC~1287                       & Cluster (Virgo)          & 1116    &  2.43  &    3.3  &   rich      & \citet{Gannon2020G}, this work   \\
NGC~5846-UDG1                  & Group (NGC~5846)         & 1014    &  1.30  &    2.1  &   rich      & \citet{Forbes2021}, this work   \\
UDG1137+16                     & Group  (Leo)             & 2167    &  1.05  &    3.3  &   unknown   & \citet{Gannon2021}, this work   \\
PUDG-R15                       & Cluster (Perseus)        & 4762    &  2.59  &    2.4  &   poor      & \citet{Gannon2022}, this work   \\   
PUDG-S74                       & Cluster (Perseus)        & 6215    &  7.85  &    3.5  &   rich      & \citet{Gannon2022}, this work   \\
PUDG-R84                       & Cluster (Perseus)        & 4039    &  2.20  &    2.0  &   rich      & \citet{Gannon2022}, this work   \\
PUDG-R24                       & Cluster$^{*}$ (Perseus)  & 7787    &  3.91  &    3.2  &   poor      & \citet{Gannon2022}, this work   \\   
PUDG-R27                       & Cluster (Perseus)        & 4376    &  4.84  &    2.1  &   rich      & this work   \\   
Yagi358 (GMP 3651)             & Cluster (Coma)          & 7969    &  1.24  &    2.3  &   rich      & \citet{Gannon2022b}, this work   \\
DFX1                           & Cluster (Coma)          & 8107    &  3.40  &    2.8  &   rich      & this work   \\
DF07                           & Cluster (Coma)          & 6600    &  4.35  &    3.8  &   rich      & this work   \\
\midrule                                                    
Literature                     &                          &         &        &         &             & \\
\midrule                                                            
Yagi093 (DF26, GMP 2748)       & Cluster (Coma)           & 6611    & 3.05   &    3.5  &   rich      & \citet{Alabi2018}, \hyperlink{FM+18}{FM+18}  \\
Yagi098                        & Cluster (Coma)           & 5980    & 1.07   &    2.9  &   rich   & \citet{Alabi2018}, \hyperlink{FM+18}{FM+18}  \\
Yagi275 (GMP 3418)              & Cluster$^{*}$ (Coma)     & 4847    & 0.94   &    2.9  &   unknown   & \citet{Alabi2018}, \hyperlink{FM+18}{FM+18}  \\
Yagi276 (DF28)                 & Cluster (Coma)           & 7343    & 1.41   &    2.3  &   unknown   & \citet{Alabi2018}, \hyperlink{FM+18}{FM+18}  \\          
Yagi392                        & Cluster (Coma)           & 7748    & 0.91   &    1.5  &   poor   & \citet{Alabi2018}, \hyperlink{FM+18}{FM+18}  \\          
Yagi418                        & Cluster (Coma)           & 8335    & 1.24   &    1.6  &   (rich)   & \citet{Alabi2018}, \hyperlink{FM+18}{FM+18}  \\  
Yagi090                        & Cluster$^{*}$ (Coma)     & 9420    & 1.00   &    2.0  &   poor   & \citet{Ruiz-Lara2018}     \\
OGS1                           & Cluster (Coma)           & 6367    & 3.10   &    1.5  &   (poor)   & \citet{Ruiz-Lara2018}     \\
DF17                           & Cluster (Coma)           & 8311    & 2.63   &    4.4  &   rich      & \citet{Gu2017}            \\
DF44                           & Cluster$^{*}$ (Coma)     & 6324    & 4.03   &    4.6  &   rich      & \citet{Villaume2022}, \citet{Webb2022}     \\
J130026.26+272735.2 (GMP~2673) & Cluster (Coma)           & 6939    & 1.56   &    3.8  &   (poor)   & \citet{Chilingarian2019}  \\
J130038.63+272835.3 (GMP~2552) & Cluster (Coma)           & 7937    & 0.64   &    1.9  &   (poor)   & \citet{Chilingarian2019}  \\
NGC~1052-DF2                   & Group (NGC~1052)         &  --     & 2.50   &    --   &   poor\textdagger& \citet{Fensch2019}        \\  
DGSAT~I                        & Field                    &  --     & 4.00   &    4.4  &   poor\textdagger& \citet{Martin-Navarro2019}\\
\bottomrule
\end{tabular}
\vspace{0.1cm}
\caption{\textbf{Main properties of the UDGs in this work:} their names (column 1); the environment they reside in (column 2): cluster, group or field, specifying to which cluster or group they belong to; their recessional velocity and stellar masses (column 3 and 4); their effective radius (column 5); and their GC classification: poor or rich (column 6) according to the number of GCs. Those within parenthesis should be taken with care (see Appendix \ref{section:ap_gcrich}). Column 7 cites the source of both the structural properties (velocities, stellar masses and sizes) and stellar populations, respectively, for the sample.\\
$^{*}$ Despite being located in clusters their local environment is considered low-density (see Section \ref{section:phasespace}).\\
\textdagger \,These objects would be considered GC-rich if the defining parameter were GC system mass instead of GC numbers, e.g. \citet{Janssens2022}.}
\end{table*}

In this paper, we study the stellar populations of this large sample of spectroscopic UDGs, to further understand the formation mechanisms that regulate the life of these intriguing galaxies. The paper is organized as follows. Section \ref{section:sample} presents the data used, for both our observations and the compilation from the literature. Section \ref{section:stellarpops} describes the methodology we use to derive the main stellar population properties of our sample, with particular attention to the different caveats of the techniques. In Section \ref{section:discussion} we discuss our results in the context of their environment UDGs reside in (Section \ref{section:phasespace}) and different relations related to other families of galaxies (Section \ref{section:stellarpopstrends}). We additionally compare to predictions for UDG stellar properties coming from simulations (Section \ref{section:simuls}). Finally, Section \ref{section:summary} presents a summary of our results. 

\begin{figure}
\centering
\includegraphics[scale=0.55]{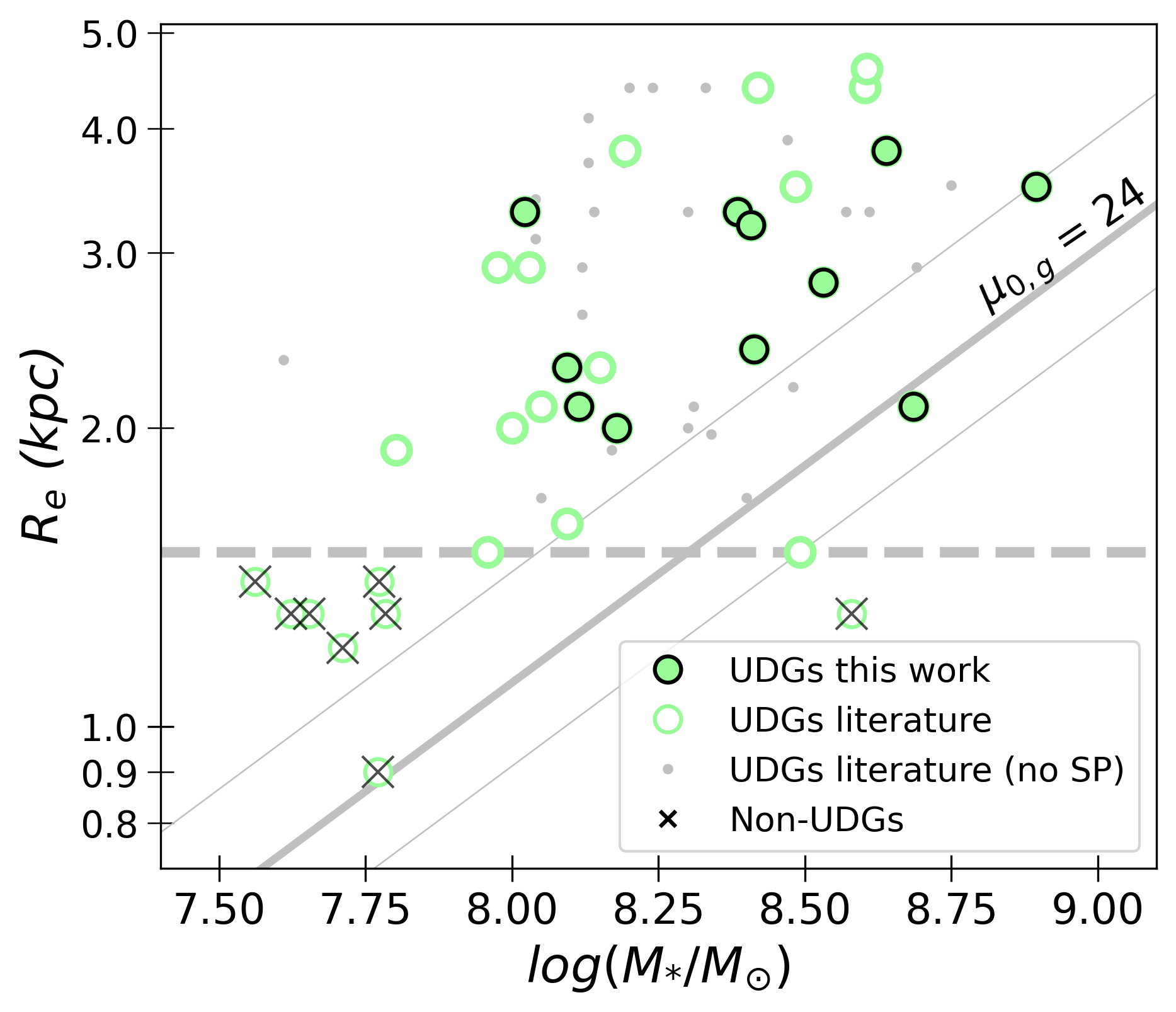}
\caption{Mass $vs.$ size diagram to separate between true UDGs and non-UDGs. The limit of \re>1.5kpc is shown by a dashed horizontal grey line, while the diagonal solid lines represent constant surface brightness limit of $\mu_{0,g}$=24 mag~arcsec$^{-2}$ with different assumptions of M/L: thickest line is for a M/L$_{g}$ of $\sim$1.6 (mean value of our sample, see Section \ref{section:agemet}), while thinner ones represent our minimum and maximum values of M/L obtained. The figure presents the UDGs in this work (filled green dots) and those from the literature sample (open green dots; data described in Section \ref{section:sample}). Other UDGs in the literature that have no spectroscopic stellar population information are shown as grey dots (\citealt{Buzzo2022}; \citealt{Gannon2022}). We highlight with an X the objects that will not be considered UDGs for this work.}
\label{fig:selection_fig}
\end{figure}

\begin{figure}
\centering
\includegraphics[scale=0.75]{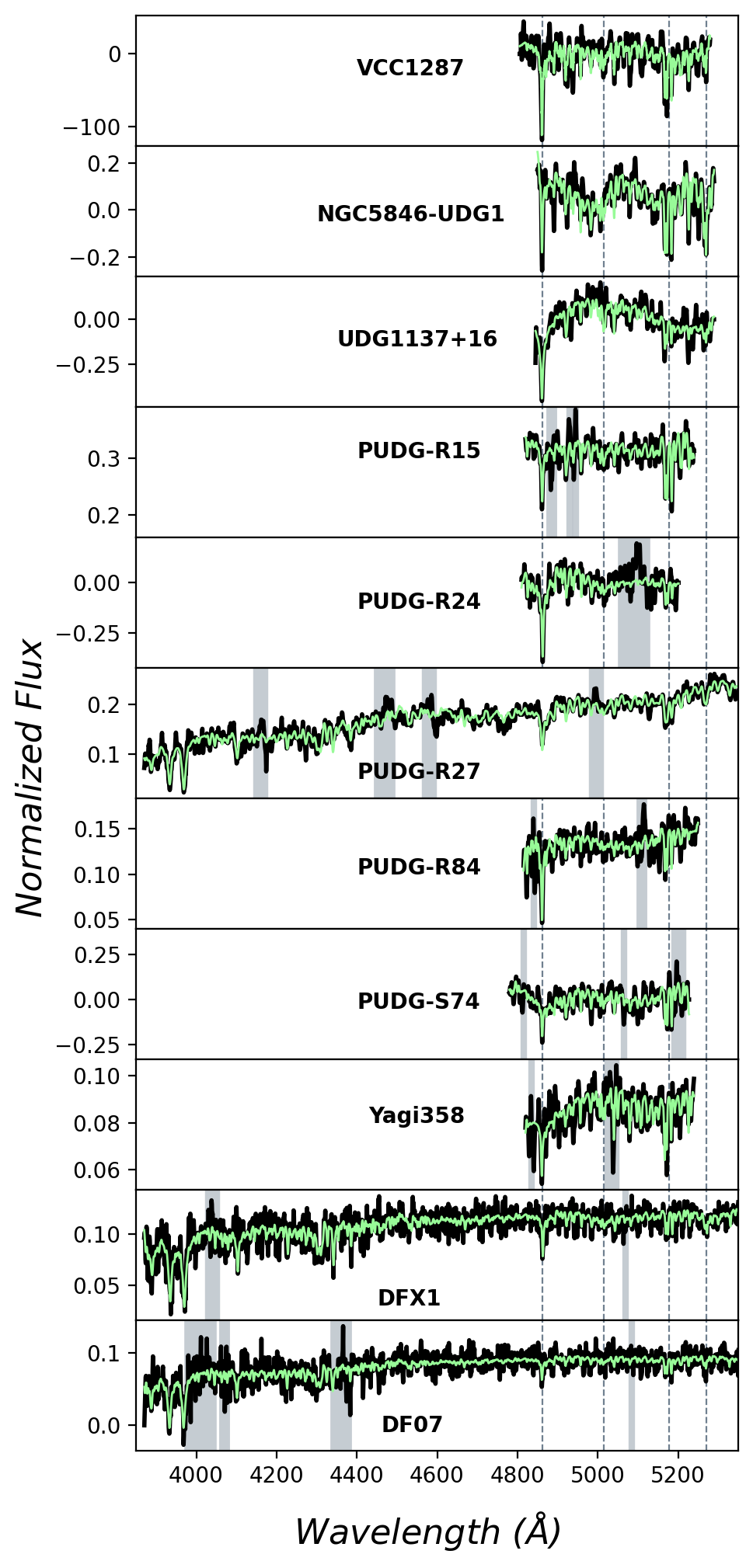}
\vspace{-0.5cm}
\caption{Spectra used in this work to derive new stellar populations for our sample of UDGs (black). The different instrumental configurations are evident: shorter wavelength and higher resolution for the BH grating, and lower resolution but larger baseline for the BL grating). The best fit from the full spectral-fitting method used in Section \ref{section:stellarpops} to derive the stellar populations is shown in green. Regions masked during the fitting process are plotted as grey vertical bands. The main line indices used in the study are shown with dashed lines (from left to right): H$_{\beta}$, Fe5015, Mg$_b$ and Fe5270.}
\label{fig:fit_fig}
\end{figure}

\section{Data}\label{section:sample}
\subsection{This work UDG sample }
In this work, we present a new stellar population analysis for a sample of quiescent UDGs with data acquired using the Keck Cosmic Web Imager (KCWI), an integral field unit in the Keck II Telescope. Details about each observation can be found in the original papers presenting the dynamical and kinematic analysis of those UDGs: the Virgo cluster UDG VCC~1287 \citep{Gannon2020G}, NGC~5846-UDG1 in the NGC~5846 group \citep{Forbes2021}, UDG1137+16 from the UGC 6594 group \citep{Gannon2021}, several UDGs in the Perseus cluster (i.e., PUDG-R15, PUDG-S74, PUDG-R84 and PUDG-R24; \citealt{Gannon2022}), and Yagi358 in the Coma cluster \citep{Gannon2022b}.  

We add to this sample three UDGs for which we have acquired new data. These are the Coma cluster UDGs DFX1 and DF07, along with the Perseus cluster UDG PUDG-R27. DFX1 and DF07 were observed on the nights 2021, April 16th and 17th (Program U105, PI: Brodie) using KCWI with the medium slicer, BL grating and a 4550 \AA~central wavelength. PUDG-R27 ($R.A.$=\,49.93116, $Dec.$=\,41.71326 in J2000) was observed on the nights of October 5th 2021 (Program W024; PI: Forbes), and January 5th and 29th 2022 (Program N195, PI: Romanowsky) using KCWI with the large slicer, BL grating and a 4550 \AA~central wavelength. Note that during the night 2022 January 29th, KCWI was slightly out of focus, which degraded the instrumental resolution by $\sim$15\%, and affected $\sim$ 20\% of our data for PUDG-R27. Data were reduced using the standard KCWI pipeline \citep{Morrissey2018} with additional cropping and flat fielding corrections as described in \citet{Gannon2020G}. For all targets, KCWI was targeted slightly off-centre to allow on-chip sky subtraction. Final exposure times were 25200 seconds for DFX1, 19200 seconds for DF07 and 26400 seconds for PUDG-R27. 

\subsection{Literature sample}
We build a literature sample using moderate S/N spectroscopic results published for the stellar population properties of Coma cluster UDGs (from \hyperlink{FM+18}{FM+18}; \citealt{Ruiz-Lara2018}; \citealt{Gu2017}; \citealt{Chilingarian2019}; and \citealt{Villaume2022}). We include the group UDG NGC~1052-DF2 \citep{Fensch2019} and the isolated UDG DGSAT~I \citep{Martin-Navarro2019}. 

Various definitions have been used to define UDGs in the literature. These are based mostly on the size and surface brightness of the objects (but see \citealt{vanNest2022} for a variety of definitions). We thus need to ensure that all the considered UDGs in this work, including those from the literature, are truly UDGs. In Figure \ref{fig:selection_fig} we select as bona-fide UDGs with \re$>1.5$~kpc (horizontal line) and $\mu_{0,g}$>24 mag~arcsec$^{-2}$ (diagonal lines, for different mass-to-light ratio, M/L, assumptions). All the UDGs from this work fulfill these criteria, with PUDG-R27 included because it lies within the surface brightness line with the highest M/L (which is in fact its own value; see Section \ref{section:agemet}). UDGs that lie below but close to the limits should not be considered true UDGs, although some might become a UDG with time (e.g., through tidal heating and surface brightness dimming; \citealt{Grishin2021}). 

Adding the sample of 14 literature UDGs to the 11 UDGs from this work we have at hand the largest spectroscopic sample of UDGs to date. Table \ref{table:bonafide} summarizes the name, environment, stellar mass, half-light radius and GC-richness of the UDGs discussed in this work. For the present work, we do not require a stringent estimate on the exact number of GCs, so we adopt a boundary between `poor' and `rich' at 20 GCs, similar to previous works (e.g., \citealt{Gannon2022}). UDGs above this limit are expected to reside in a relatively massive dark matter halo under the assumption that they obey the GC number -- halo mass relationship of \citet{Burkert2020}. 

The GC richness estimates are collected from a variety of sources in the literature (e.g., \citealt{Forbes2020,Danieli2022,Gannon2022}). There are 10 UDGs in our sample without this information, all from the Coma cluster. For some of these we have carried out new estimates of their GC richness, as described in Appendix~\ref{section:ap_gcrich}, following a similar visual approach to \citet{Gannon2022}. These estimates are primarily based on {\it HST} imaging, and in some cases on deep ground-based imaging from Subaru/Suprime-Cam.

Figure \ref{fig:fit_fig} presents the spectra for the UDGs whose stellar populations are newly presented in this work (black). Different KCWI instrumental configurations were used. The medium slicer with a BH3 grating provides a shorter baseline ($\sim$4800--5200\AA) but higher spectral resolution (0.5\AA\, FWHM), crucial to derive proper velocity dispersion measurements. The BL grating provides a longer baseline ($\sim$3800--5400\AA) but with lower spectral resolution (5.1\AA\, FWHM), adequate for stellar population studies. PUDG-R84 was observed in both configurations as a consistency check. However, for this UDG we will use the results from the BH configuration, as it has a much higher S/N than the BL one. One caveat to have in mind for the UDGs observed with the BH grating is that a short wavelength range can introduce systematics on the derived properties. Shorter baselines (in particular when not including the age indicator H$_{\beta}$) tend to give younger ages (see e.g., \citealt{Forbes2022}). Nonetheless, as later discussed in Appendix \ref{section:ap_agemet}, this is not affecting our results. 

\section{Stellar population analysis}\label{section:stellarpops}
Before proceeding, we need to highlight a caveat in the data. Some of the spectra have been reduced to have a pedestal removed from their continuum as a result of the sky subtraction process. This is due to an inability of the PCA sky subtraction routine described in \citet{Gannon2020G} to disentangle the absolute level of the continuum coming from the galaxy and the sky emission. This results in those spectra having negative counts at some wavelengths (see Fig.~\ref{fig:fit_fig}). While this does not have any impact on the full spectral fitting procedure detailed below, it means that we can not perform a line index measurement for VCC~1287, NGC~5846-UDG1, UDG1137+16, PUDG-R24 and PUDG-S74.

We therefore use both the full spectral fitting code \ppxf \citep{Cappellari2017)} and perform a classical absorption line index measurement (when possible) to our sample of UDGs to obtain mean mass-weighted ages, metallicities, and $\alpha$-enhancement ratios. We also derive the SFHs, from which one can calculate crucial timescales such as the quenching timescale. We employ the MILES single stellar population (SSP) models \citep{Vazdekis2015} assuming a Kroupa initial mass function \citep{kroupa2001}. These models range in age from 0.03 to 14 Gyr and metallicities of [M/H] = $-$2.42 to $+$0.40 dex. We use the scaled-solar base models to obtain the stellar population parameters such as ages, metallicities and SFHs. 

First, the optimal degree for the polynomials to be used in the fitting is found for each UDG, which is typically of the order of 5 for both the additive and multiplicative polynomials (see for example \citealt{Dago2023} for a detailed discussion on how to determine these values). \ppxf is then run to obtain a non-regularized solution (it will be the closest to a line index approach, a discrete SFH), and also a regularized one. The latter provides a more realistic SFH while maintaining an accurate fit given its simple Bayesian approach (see e.g., \citealt{Cappellari2017)}; \citealt{McDermid2015} and \citealt{Westfall2019} for additional information on this process). We have checked that the results are independent of the regularization choice, showing similar SFHs and stellar populations properties. We therefore use the average between the regularized and un-regularized solution for the ages, metallicities and timescales presented here. All uncertainties associated with the stellar population values are calculated by performing Monte Carlo simulations of this fitting procedure for each UDG. 

\subsection{$\alpha$-enhancement ratios}\label{section:alphas}
The $\alpha$-enhancement is a challenging measurement that has a high dependency on the methodology used. For example, if implemented in a full spectral fitting routine as another free parameter (together with the age and metallicity), the values obtained will only span the values of the SSP models used. In the case of the models used here, those values will be between \afe= 0.0 and $+$0.4\,dex. Instead, if calculated using line index fitting, there are no limitations besides how far one can extrapolate the values outside the model grids. With this last method one obtains [Mg/Fe], as those are the line indices typically used. One can convert from one to the other (\afe=0.02+0.56$\times$[Mg/Fe], if using the BaSTI Base models; \citealt{Vazdekis2015}). In addition, the \afe\, obtained from the full spectral fitting approach can be underestimated for galaxies with low metallicities. This is because the SSP models are based on solar-scaled stars, which in reality are Mg enhanced at the low metallicity regime ($\lesssim\,-$1\,dex; \citealt{Vazdekis2015}), i.e. similar to the regime covered by UDGs.  

\begin{figure}
\centering
\includegraphics[scale=0.55]{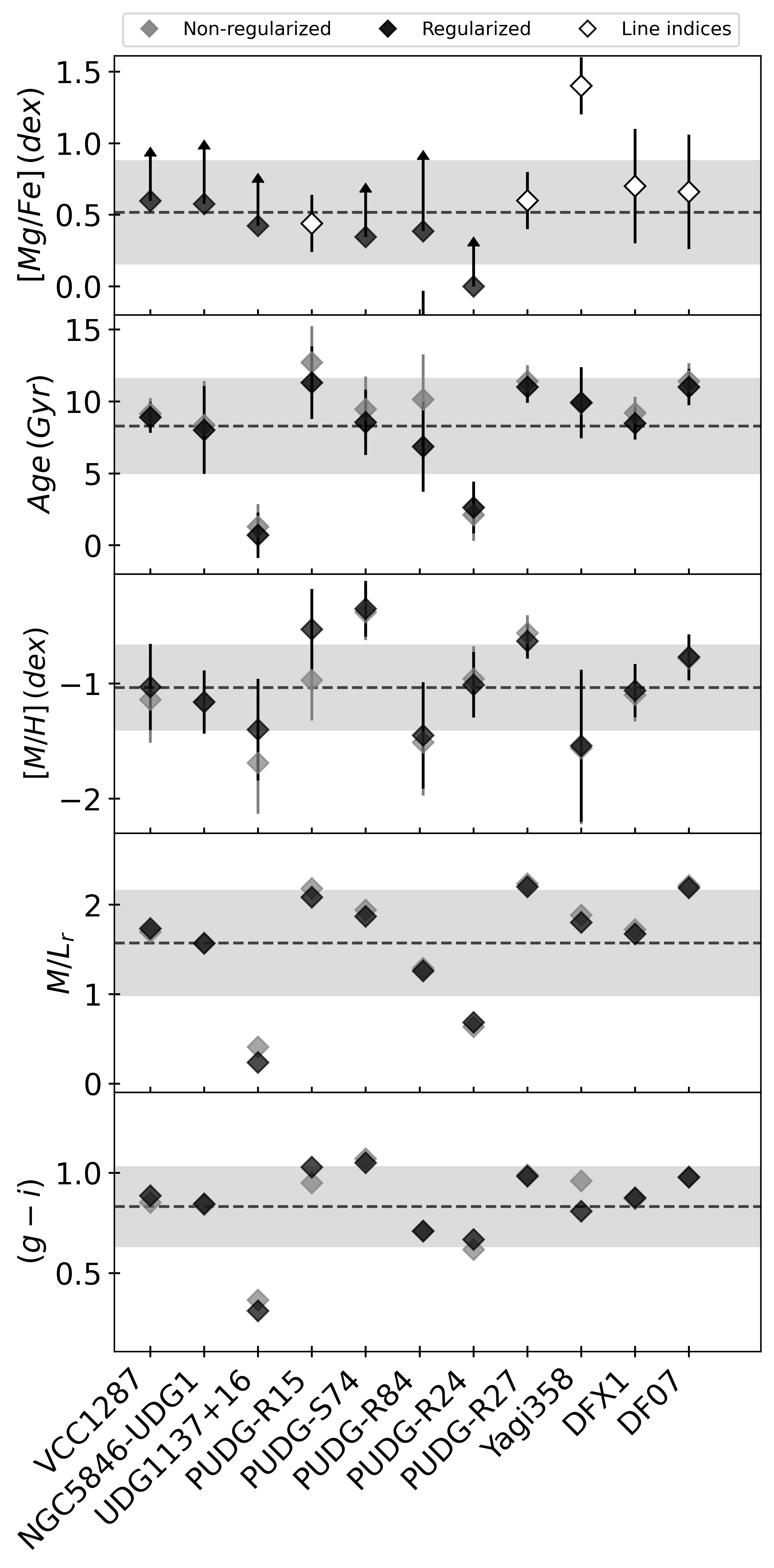}
\caption{Stellar populations for the UDGs in this work: [Mg/Fe], mass-weighted age, mass-weighted metallicity, the derived M/L in the $r$-band, and the $g-i$ colour (from top to bottom, respectively). For each property we present the individual measurements for each UDG (diamonds) and the average value for all UDGs (dashed line) with the 1-$\sigma$ deviations (grey band). Grey symbols correspond to the non-regularized solution while black ones correspond to the regularized one from the full spectral fitting with \ppxf. White symbols in the top panel correspond to line index measurements (see Appendix \ref{section:ap_alfas} for a discussion on the different ways to obtain the [Mg/Fe]).}
\label{fig:pops_fig}
\end{figure}

Taking all this into consideration, it is difficult to unify the ways different works measure the ratio between $\alpha$-elements and iron. Appendix \ref{section:ap_alfas} contains a section fully dedicated to the different ways of calculating it and their systematics. Unfortunately we can not use this more robust line index measurement for 5 of our UDGs due to the pedestal issue discussed in Section \ref{section:stellarpops}. We will therefore use, when available, the [Mg/Fe] value from the line index measurement. Otherwise, we will use the one from \ppxf, cautioning that these may be lower limits. Moreover, we note that the abundance ratios obtained from the line-index measurement are typically above the averaged value and higher than the SSP limit marked by the full spectral fitting approach. This highlights the importance of being careful about the way one computes this value. The results from line index measurements are marked with a \textdagger\ in Table \ref{table:ssp}, and are shown as an open symbol in the top panel of Figure \ref{fig:pops_fig}.

We find that the majority of our UDGs present elevated [Mg/Fe] abundance patterns, with a mean value and dispersion of $\langle$[Mg/Fe]$\rangle=+0.51\pm0.32$\,dex, slightly higher than what was found in \hyperlink{FM+18}{FM+18}. Converting this [Mg/Fe] value into \afe\, would result in an enhanced value of \afe$\sim$0.32\,dex. PUDG-R24 is the only UDG that shows a scaled-solar pattern.

\subsection{Mean ages and metallicities}\label{section:agemet}
We obtain several parameters from the \ppxf fitting such as the mean ages (t$_M$, in lookback time), mean metallicities ([M/H]), the stellar mass-to-light ratios ($M/L$ in the $r$-band) and the predicted $g-i$ colour. 
The resulting stellar populations properties are shown in Figure \ref{fig:pops_fig}. For each, we present the mass-weighted results for both the regularized (black symbols) and non-regularized (grey symbols) solutions. Average values between regularised and non-regularised results were obtained and these are the ones that will be used for the remainder of this paper. These are summarized in Table \ref{table:ssp}. 

The mean mass-weighted age of our sample of UDGs is $\langle \,t_\mathrm{M}\,\rangle$=8.3$\pm$3.3\,Gyr. Three UDGs have very old ($t_\mathrm{M}$>10~Gyr) mass-weighted ages, six UDGs have old ages (between 7.5 and 10\,Gyr), and two UDGs are young (<4.5\,Gyr). Due to these large age differences, the M/L values of their stellar populations vary from roughly 0.3 to 2. We remind the reader of the caveat that shorter wavelength ranges may deliver younger ages, discussed in Appendix \ref{section:ap_agemet}. In this work, this systematic bias may have had an effect on the two young UDGs. However, PUDG-R24 has been identified as a very recent infall into the Perseus cluster and UDG1137+16 presents tidal features, which can in both cases explain the young stellar ages. 

The UDGs in this work all have sub-solar metallicities, with a mean value of $\langle$[M/H]$\rangle=-1.03\pm$0.37\,dex, similar to previous works (e.g. \hyperlink{FM+18}{FM+18}; \citealt{Gu2017}; \citealt{Ruiz-Lara2018}). The mean colour of the UDGs, $\langle g-i \rangle=0.8\pm0.2$, is compatible with the one found in other large samples of UDGs (e.g., \citealt{Roman2017a}; \citealt{Junais2022}). 

We note that several of the UDGs in the sample already have stellar populations results available. VCC~1287, PUDG-R24, Yagi358, DFX1 and DF07 have published SED fitting stellar populations (e.g., \citealt{Pandya2017}; \citealt{Buzzo2022}). As mentioned above, here we only consider spectroscopic results, but we refer the reader to Appendix \ref{section:ap_complit} for a comparison to these SED fitting results. While stellar masses and ages do not differ significantly from the SED solution, stellar metallicities from SED fitting are systematically lower, on the order of $\sim$0.25\,dex. NGC~5846-UDG1 was previously studied in \citet{Mueller2020} using MUSE spectroscopy with a slightly lower S/N than our new data (see also \citealt{Heesters2023}), and DF07 in \citet{Gu2017} from MaNGA data. These works reported a [Fe/H]=--1.33$\pm$0.19\,dex and an age of 11.2$\pm$1.5 for NGC~5846-UDG1 and [Fe/H]=--1.03$\pm$0.31\,dex and an age of 9$\pm$1.5\,Gyr for DF07. These are compatible with our measurements within the uncertainties. Therefore, we will use our own values for the remaining of the paper.

\subsection{Star formation histories and characteristic timescales}\label{section:sfhs}
From the recovered SFHs, different characteristic lookback times are calculated: when the galaxy formed half of its stellar mass (t$_{50}$), and when it built 90\% of it (t$_{90}$). The latter is also considered to be a good estimate for the quenching time of the galaxy (e.g., \citealt{Weisz2014b}; \citealt{Weisz2019}; \citealt{Collins2022}). These look-back times are translated into timescales (i.e., the number of years it took to build that given amount of stellar mass). This way, we calculate $\Delta$t$_{50}$= t$_\mathrm{BB}$–t$_{50}$ and $\Delta$t$_{90}$= t$_\mathrm{BB}$–t$_{90}$, where t$_\mathrm{BB}$is the time since the Big Bang, 13.8\,Gyr. Therefore we are using here as a quenching timescale the commonly used timescale \textit{since the Big Bang}, not since the onset of star formation.

Under the assumption that quenching is environmentally driven, it is thought that galaxies quench roughly 1--2\,Gyr after falling into the cluster or group environments (e.g., \citealt{Muzzin2008}: \citealt{Fillingham2015}). We thus estimate a time of infall as 1.5 Gyr before quenching, i.e.,  t$_\mathrm{inf}\sim$t$_{90}-1.5$\,Gyr (see also \hyperlink{FM+18}{FM+18}). Table \ref{table:ssp} also quotes these relevant timescales inferred from the SFHs. 

\begin{table*}
\centering
\label{table:ssp}                      
\begin{tabular}{c |c c c c c c c c}   
\toprule      
\textbf{Galaxy}     &  Age ($t_{M}$)   &   t$_{50}$         &     t$_{90}$   & t$_\mathrm{inf}$  & $\Delta$t$_{50}$ & $\Delta$t$_{90}$  & [M/H]                 & [Mg/Fe] \\
                    & (Gyr)            &   (Gyr)            &     (Gyr)      &   (Gyr)                 & (Gyr)            &    (Gyr)          & (dex)                 & (dex)\\
\midrule            
\midrule            
 VCC~1287           &   9.09$\pm$ 1.07 &    9.2 $\pm$ 0.2   &  8.5 $\pm$ 1.1 &   10.0 $\pm$ 1.1        &   4.7 $\pm$ 0.2  &   5.9 $\pm$ 1.1   &  $-$1.06 $\pm$ 0.34  &  0.56 $\pm$ 0.11  \\
 NGC~5846-UDG1      &   8.20$\pm$ 3.05 &    8.3 $\pm$ 0.1   &  7.5 $\pm$ 1.0 &     --                  &   5.5 $\pm$ 0.1  &   6.8 $\pm$ 1.0   &  $-$1.15 $\pm$ 0.25  &  0.54 $\pm$ 0.18  \\
 UDG1137+16         &   2.13$\pm$ 1.58 &    1.0 $\pm$ 0.3   &  0.1 $\pm$ 0.1 &     --                  &  12.9 $\pm$ 0.3  &  13.7 $\pm$ 0.1   &  $-$1.52 $\pm$ 0.40  &  0.39 $\pm$ 0.10  \\
 PUDG-R15           &  11.32$\pm$ 2.52 &   13.4 $\pm$ 0.8   &  9.5 $\pm$ 0.1 &   11.0 $\pm$ 0.1        &   0.8 $\pm$ 0.8  &   4.5 $\pm$ 0.1   &  $-$0.93 $\pm$ 0.32  &  0.44 $\pm$ 0.20\textdagger  \\
 PUDG-S74           &   8.44$\pm$ 2.26 &    8.2 $\pm$ 3.3   &  9.5 $\pm$ 0.1 &   11.0 $\pm$ 0.1        &   4.2 $\pm$ 3.3  &   7.3 $\pm$ 0.1   &  $-$0.40 $\pm$ 0.22  &  0.32 $\pm$ 0.11  \\
 PUDG-R84           &   8.99$\pm$ 3.20 &    8.4 $\pm$ 0.5   &  6.8 $\pm$ 0.4 &    8.5 $\pm$ 0.4        &   5.7 $\pm$ 0.5  &   7.2 $\pm$ 0.4   &  $-$1.48 $\pm$ 0.46  &  0.22 $\pm$ 0.30  \\
 PUDG-R24           &   2.58$\pm$ 1.80 &    3.1 $\pm$ 0.1   &  0.8 $\pm$ 0.1 &    2.5 $\pm$ 0.1        &  10.5 $\pm$ 0.1  &  12.9 $\pm$ 0.1   &  $-$1.00 $\pm$ 0.26  &  0.03 $\pm$ 0.08  \\
 PUDG-R27           &  11.04$\pm$ 1.11 &   10.3 $\pm$ 0.6   &  8.8 $\pm$ 0.7 &   10.0 $\pm$ 0.7        &   3.1 $\pm$ 0.6  &   5.3 $\pm$ 0.7   &  $-$0.61 $\pm$ 0.14  &  0.31 $\pm$ 0.20\textdagger  \\
 Yagi358            &   9.81$\pm$ 2.46 &   13.2 $\pm$ 1.3   &  6.1 $\pm$ 0.1 &    7.5 $\pm$ 0.1        &   1.3 $\pm$ 1.3  &   7.4 $\pm$ 0.1   &  $-$1.56 $\pm$ 0.60  &  1.40 $\pm$ 0.20\textdagger  \\
 DFX1               &   8.84$\pm$ 1.13 &    9.1 $\pm$ 0.8   &  8.1 $\pm$ 1.4 &    9.5 $\pm$ 1.4        &   5.1 $\pm$ 0.8  &   6.5 $\pm$ 1.4   &  $-$1.08 $\pm$ 0.21  &  0.70 $\pm$ 0.40\textdagger  \\
 DF07               &  11.18$\pm$ 1.27 &   10.5 $\pm$ 1.0   &  9.1 $\pm$ 0.7 &   10.5 $\pm$ 0.7        &   2.8 $\pm$ 1.0  &   5.0 $\pm$ 0.7   &  $-$0.78 $\pm$ 0.18  &  0.68 $\pm$ 0.40\textdagger  \\
 \midrule                                                                                                                                     
Yagi093             &   7.88$\pm$ 1.76 &    11.3    &  4.1 $\pm$ 0.1 &    5.5 $\pm$ 0.1        &   3.2   &  10.0 $\pm$ 0.1   &  $-$0.56 $\pm$ 0.18  &   0.38$\pm$ 0.17 \\
Yagi098             &   6.72$\pm$ 2.16 &    11.3    &  2.1 $\pm$ 2.9 &    3.5 $\pm$ 2.9        &   2.8   &  10.5 $\pm$ 2.9   &  $-$0.72 $\pm$ 0.20  &    --            \\
Yagi275             &   4.63$\pm$ 1.50 &     5.1    &  2.2 $\pm$ 0.1 &    4.0 $\pm$ 0.1        &   4.7   &  11.8 $\pm$ 0.1   &  $-$0.37 $\pm$ 0.17  &  $-$0.25$\pm$ 0.38 \\
Yagi276             &   4.24$\pm$ 2.32 &     9.0    &  2.0 $\pm$ 0.1 &    3.5 $\pm$ 0.1        &   4.9   &  12.0 $\pm$ 0.1   &  $-$0.38 $\pm$ 0.79  &    --            \\
Yagi392             &   7.36$\pm$ 2.06 &    10.5    &  2.8 $\pm$ 2.7 &    4.5 $\pm$ 2.7        &   4.3   &   8.2 $\pm$ 2.7   &  $-$0.58 $\pm$ 0.73  &    --            \\
Yagi418             &   7.87$\pm$ 2.02 &    12.3    &  3.2 $\pm$ 2.5 &    4.5 $\pm$ 2.5        &   4.0   &   9.9 $\pm$ 2.5   &  $-$1.10 $\pm$ 0.85  &   0.17$\pm$ 0.31 \\
Yagi090             &   5.75$\pm$ 1.30 &     9.4    &  5.0 $\pm$ 0.1 &    6.5 $\pm$ 0.1        &   4.4   &   9.6 $\pm$ 0.1   &  $-$1.35 $\pm$ 0.05  &   0.40  \\
OGS1                &   8.50$\pm$ 1.20 &    13.8    & 10.4 $\pm$ 0.7 &   11.5 $\pm$ 0.7        &   0.1   &   3.3 $\pm$ 0.7   &  $-$0.53 $\pm$ 0.06  &   0.35  \\
DF17                &   9.11$\pm$ 2.00 &     --     &        --      &          --             &   --    &       --          &  $-$0.83 $\pm$ 0.50  &   --             \\
DF44                &  10.23$\pm$ 1.50 &    0.34    &  3.0 $\pm$ 4.7 &    4.5 $\pm$ 4.7        &   2.8   &   6.3 $\pm$ 4.7   &  $-$1.33 $\pm$ 0.05  &  $-$0.10$\pm$ 0.10 \\
J130026.26+272735.2 &    1.5$\pm$ 0.10 &     --     &        --      &          --             &   --    &       --          &  $-$1.04 $\pm$ 0.11  &   --             \\
J130038.63+272835.3 &    1.7$\pm$ 0.10 &     --     &        --      &          --             &   --    &       --          &  $-$0.74 $\pm$ 0.08  &   --             \\
NGC~1052-DF2        &   8.90$\pm$ 1.50 &     --     &        --      &          --             &   --    &       --          &  $-$1.07 $\pm$ 0.11  &   0.10$\pm$ 0.05 \\
DGSAT~I             &   8.10$\pm$ 0.40 &     --     &        --      &          --             &   5.6   &      13.5         &  $-$1.80 $\pm$ 0.40  &   1.50$\pm$ 0.50 \\
\bottomrule
\end{tabular}
\vspace{0.1cm}
\caption{\textbf{Stellar population properties of the UDGs in this work (top block) and literature (bottom)}. For this sample, the values correspond to a median between the non-regularized and regularized solution for the mean mass-weighted age (column 2); t$_{50}$ and t$_{90}$ (lookback times when the galaxy formed 50 and 90\% of its stellar mass; columns 3 and 4, respectively). Those UDGs for which this value is not stated in the original works, but the SFH is shown, have been digitized to obtain these values. t$_\mathrm{inf}$ is an estimate of the expected lookback time of cluster infall assuming quenching is due to cluster/group infall (t$_\mathrm{inf}$=t$_{90}-$1.5\,Gyr, column 8). $\Delta$t$_{50}$ and $\Delta$t$_{90}$ are the timescales it took to build 50 and 90\% of stellar mass since the Big Bang (columns 6 and 7, respectively). } Column 8 shows the mean mass-weighted metallicity and column 9 the [Mg/Fe], with \textdagger\, being the values obtained from the line index approach (see Appendix \ref{section:ap_alfas}). We note that \afe\, values from the literature have been transformed into [Mg/Fe] using the corresponding conversion from \citet{Vazdekis2015}. 
\end{table*}

\section{Discussion}\label{section:discussion} 
\subsection{SFHs and their relation with environment}\label{section:envGC}
The current galaxy formation paradigm predicts that the mass ratio between $\alpha$ elements and iron is a good proxy for the formation timescales of galaxies (e.g., \citealt{Matteucci1994}; \citealt{Thomas2005}; \citealt{deLaRosa2011}; \citealt{McDermid2015}; \textbf{\citealt{Romero-Gomez2023})}. Typically, galaxies with younger ages (from more extended SFHs) tend to have near solar values, while those with very fast formation tend to have elevated abundance ratios. However, one of the most puzzling results to date for UDGs was that overall they show elevated $\alpha$-enhancement, despite presenting, in many cases, extended SFHs and intermediate ages (e.g., \hyperlink{FM+18}{FM+18}; \citealt{Martin-Navarro2019}; \citealt{Rong2020a}). 

\begin{figure*}
\centering
\includegraphics[scale=0.48]{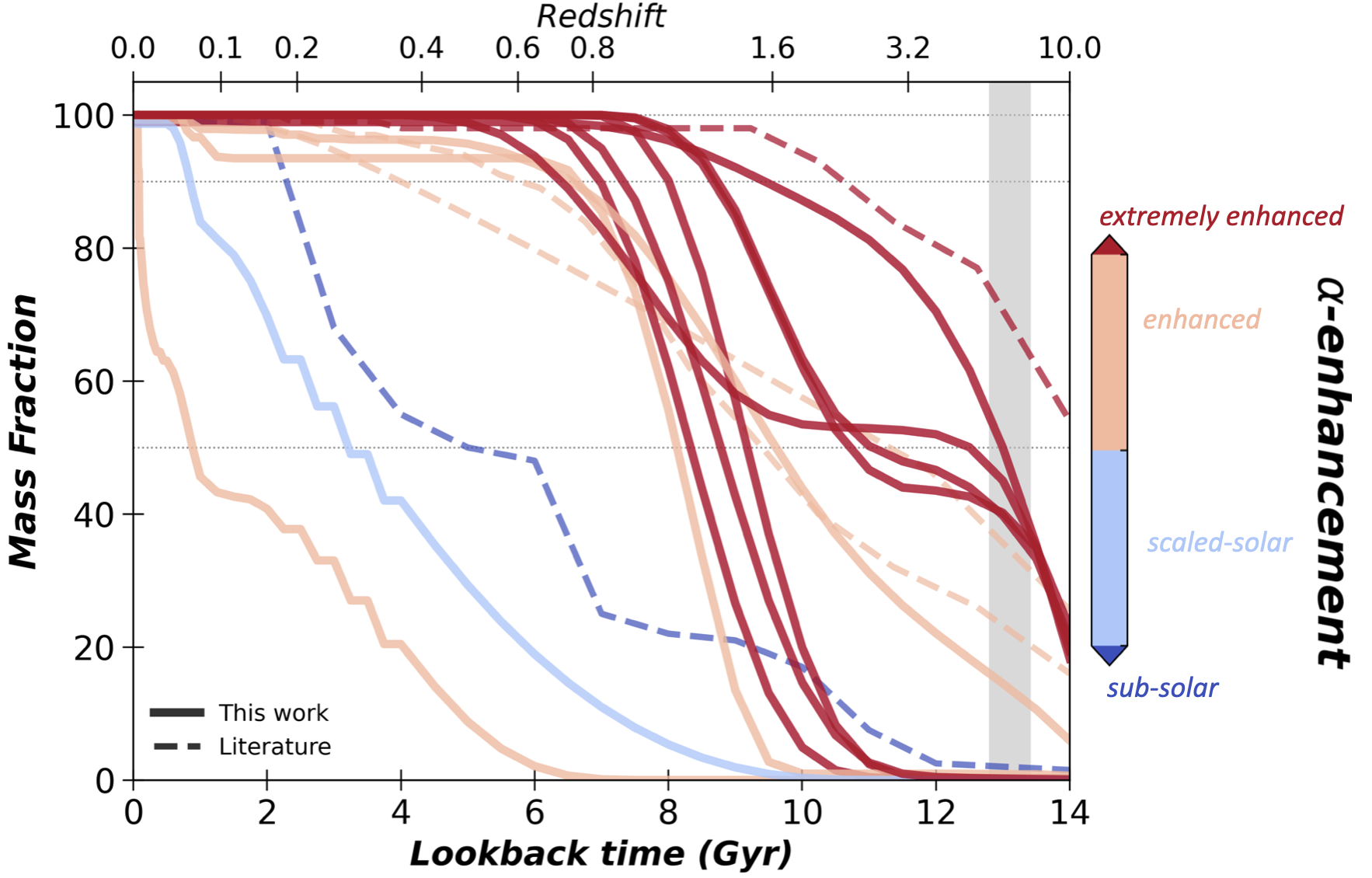}
\caption{Cumulative mass fraction of each UDG in this work (solid line), colour-coded by their level of $\alpha$-enhancement. We note that this color classification is consistent regardless of using \afe\, or [Mg/Fe], and has been chosen to unify the different results from the techniques presented in Appendix \ref{section:ap_alfas}. We also include other Coma UDGs from FM+18 and \citet{Ruiz-Lara2018} with existing [Mg/Fe] abundance ratios (dashed lines). Horizontal dotted lines mark the 50, 90 and 100\% mass fraction in stars. We note that none of the UDGs was quenched by reionization, marked as a grey vertical band \citep{Fan2006}. A variety of SFHs are found for UDGs, with a clear relation with the $\alpha$-enhancement. Rapid SFHs tend to show elevated values regardless of whether or not the galaxy started forming very early on (extremely high, >0.4\,dex) or delayed in time (high, 0.2--0.4\,dex). In contrast, those with more extended SFHs tend to be scaled solar (<0.2\,dex) or even sub-solar (<0.0\,dex). }
\label{fig:sfhs_all_fig}
\end{figure*}

\begin{figure*}
\centering
\includegraphics[scale=0.78]{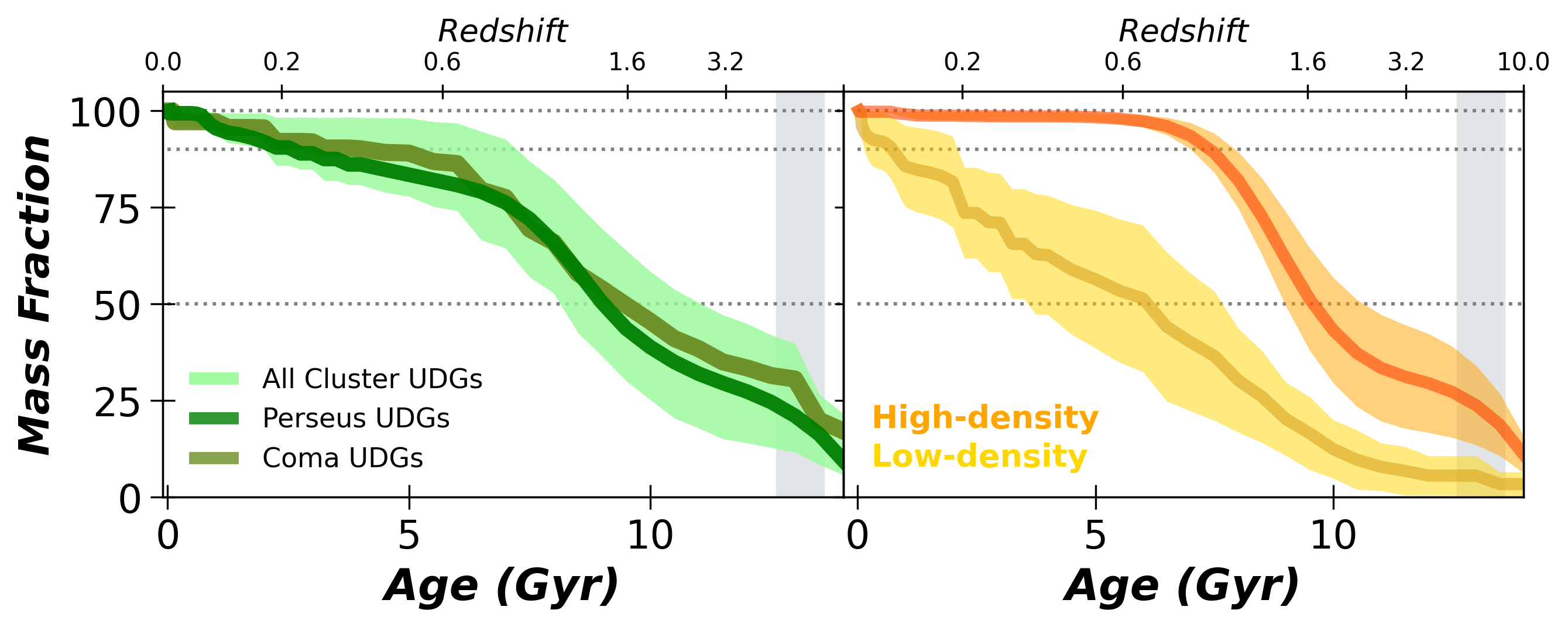}
\vspace{-0.2cm}
\caption{Similar to Figure \ref{fig:sfhs_all_fig} but now UDGs are averaged by different properties. \textit{Left panel:} The average of all cluster UDGs from this work, regardless of their local environment, with the intrinsic scatter shown as the shaded light green area. Perseus and Coma UDG averages are show by solid green lines as described in the legend. It seems that UDGs build up their stellar mass in a similar way in the Perseus and Coma clusters. \textit{Right panel:} all UDGs in this work separated by their local environment: low-density regions such as field, groups, and recent infalls (yellow), and high-density regions as the rest of cluster UDGs (orange). We find that UDGs in high-density environments tend to form earlier and faster than those in low-density environments. }
\label{fig:sfh_byprop}
\end{figure*}

To further investigate the relation between the SFHs and the $\alpha$-enhancement, Figure \ref{fig:sfhs_all_fig} presents the regularized SFHs as the cumulative mass fraction over cosmic time of the 11 UDGs in this work and other Coma UDGs from the literature for which we have access to (\hyperlink{FM+18}{FM+18} and \citealt{Ruiz-Lara2018}). They are colour-coded by their level of $\alpha$-enhancement: sub-solar (<0\,dex), scaled-solar (0 to +0.2\,dex), enhanced (+0.2\,dex to +0.4\,dex), and extremely enhanced (>0.4\,dex). This has been done to unify measurements from different techniques and works (e.g. the use of \afe\, or [Mg/Fe]), as we are here interested in the level of enhancement rather than the absolute value of it. 

Figure \ref{fig:sfhs_all_fig} shows distinctly different types of SFHs. While most of the UDGs are quenched on relatively short timescales (around 8\,Gyr ago), none of them was quenched during the epoch of Reionization, marked with a grey band in the figure \citep{Fan2006}. A clear trend with the $\alpha$-enhancement is seen. Some UDGs started forming stars very early on and built their entire mass relatively fast (\textit{early-and-fast}). Hence they show very short quenching timescales ($\Delta$t$_{90}\sim$4--5\,Gyr), thus expected to be early cluster infallers, and have the oldest stellar ages ($t_\mathrm{M}\sim$10--13Gyr). We note that galaxies with similar SFHs are found in the Local Group (e.g., such as And~XIX; \citealt{Collins2022}). Moreover, these early-and-fast UDGs also show the highest $\alpha$-enhancement values (> 0.4\,dex). 

Similarly high $\alpha$-enhancements are seen for several UDGs that present a fast build-up but that their onset of star formation occurred later in cosmic time ($\sim$4\,Gyr after the Big Bang). This means that while the quenching timescales of these \textit{late-and-fast} UDGs are as short as to the early-and-fast type, they will present intermediate stellar ages ($t_\mathrm{M}\sim$6--9\,Gyr). 

Some other UDGs also started forming very early on, however they did not quench for a much longer time (\textit{early-and-slow}), sometimes even later than the late-and-fast UDGs. These UDGs, also mainly located in clusters, show intermediate ages due to their more extended SFHs, and their $\alpha$-enhancement, despite being lower than in the previous two types, are still relatively high ($\sim$0.2--0.4\,dex). 

Finally, there are three UDGs that show very late, extended SFHs, quenching only 1--3\,Gyr ago (\textit{late-and-slow}). They present extremely young stellar ages, suggestive of a late cluster infall. From these late-and-slow UDGs, the two located in clusters are the ones with the lowest $\alpha$-enhancements ($\lesssim$ 0.2\,dex; PUDG-R24 and Yagi275). Such extremely late SFHs have also been reported for other classical dwarf galaxies (see e.g., \citealt{Weisz2014b}; \citealt{Ruiz-Lara2018b}; \citealt{Weisz2019}; \citealt{Bidaran2022}). In \citealt{Bidaran2022} it was suggested that these recent bursts of star formation can be associated to late cluster infall. This is compatible for the two cluster UDGs, while the remaining one, the group UDG1137+16, presents a slightly higher abundance ratio than its cluster counterparts, despite sharing a similar SFH type. However, this is a UDG with tidal features and therefore this could be the result of a past interaction.  

There seems to exist a clear relation between the SFHs, the ages and the $\alpha$-enhancement of UDGs, which in turn is related to the environment they reside in. Moreover, there seems to be a trend with stellar mass, with those UDGs quenching the fastest (and therefore with the most elevated [Mg/Fe]), being all on the massive end of the UDG population. To further investigate this, Figure \ref{fig:sfh_byprop} shows the result of averaging the SFHs of the UDGs according to their environment. As the majority of the UDGs in our sample and in the literature are located in clusters, the left panel focuses on this environment alone. The grey region shows the average SFH and the intrinsic scatter for all cluster UDGs (from this work, \hyperlink{FM+18}{FM+18} and \citealt{Ruiz-Lara2018}). On average, cluster UDGs start forming early, reaching 50\% of their stellar masses in $\sim$4--5~Gyr after the Big Bang. Both Coma and Perseus UDGs seem to build up their stellar mass in a similar way, although Coma UDGs quench slightly earlier than Perseus ones (5 vs. 2\,Gyr, respectively). However, comparing these averages with the individual SFHs in \ref{fig:sfhs_all_fig} it is clear that they are not a representative of the whole cluster population, as some UDGs have clearly much earlier and faster SFHs.  

If UDGs fell into the cluster late, this will skew the results toward younger, more extended SFHs. This is the case of PUDG-R24, Yagi275 from \hyperlink{FM+18}{FM+18} and Yagi90 from \citet{Ruiz-Lara2018} (see also Section \ref{section:phasespace}). In such cases, their immediate environment is better represented by a low-density one, more similar to groups. The right panel of Figure \ref{fig:sfh_byprop} now separates our complete sample of UDGs by (local) environment: low-density (field, group and cluster outskirts) and high-density (the rest of cluster UDGs). With this separation we find that UDGs in high-density environments formed earlier and faster, having built the bulk of their stars $\sim$8-9\,Gyr ago. Instead, UDGs in low-density regions on average started forming with a $\sim$4\,Gyr delay, forming stars until almost the present day. This result supports the trends found by \citet{Buzzo2022} and \citet{Barbosa2020}, where older UDGs tend to be found in denser environments than younger ones. We caution, however, that low-density UDGs conform about half of the UDGs with SFHs, and that for example there are no field galaxies in this group of UDGs. 

Similar SFH trends have been found in Local Group dwarfs (e.g. \citealt{Gallart2015}; \citealt{Weisz2015}; \citealt{Ruiz-Lara2018b}; \citealt{Collins2022}). In fact, \citet{Gallart2015} separated these dwarf galaxies between fast (if quenched around 8 Gyr ago) and slow (that started forming later in time and that show more extended SFHs). Interestingly, if we split our UDGs into fast/slow according to the timescales derived above, we find virtually the same SFHs (same shape and formation timescales) for the fast and high-density UDGs, and for the slow and low-density ones. This further reinforces the result that environment seems to play a role in how UDGs form. 

\begin{figure*}
\centering
\includegraphics[scale=0.75]{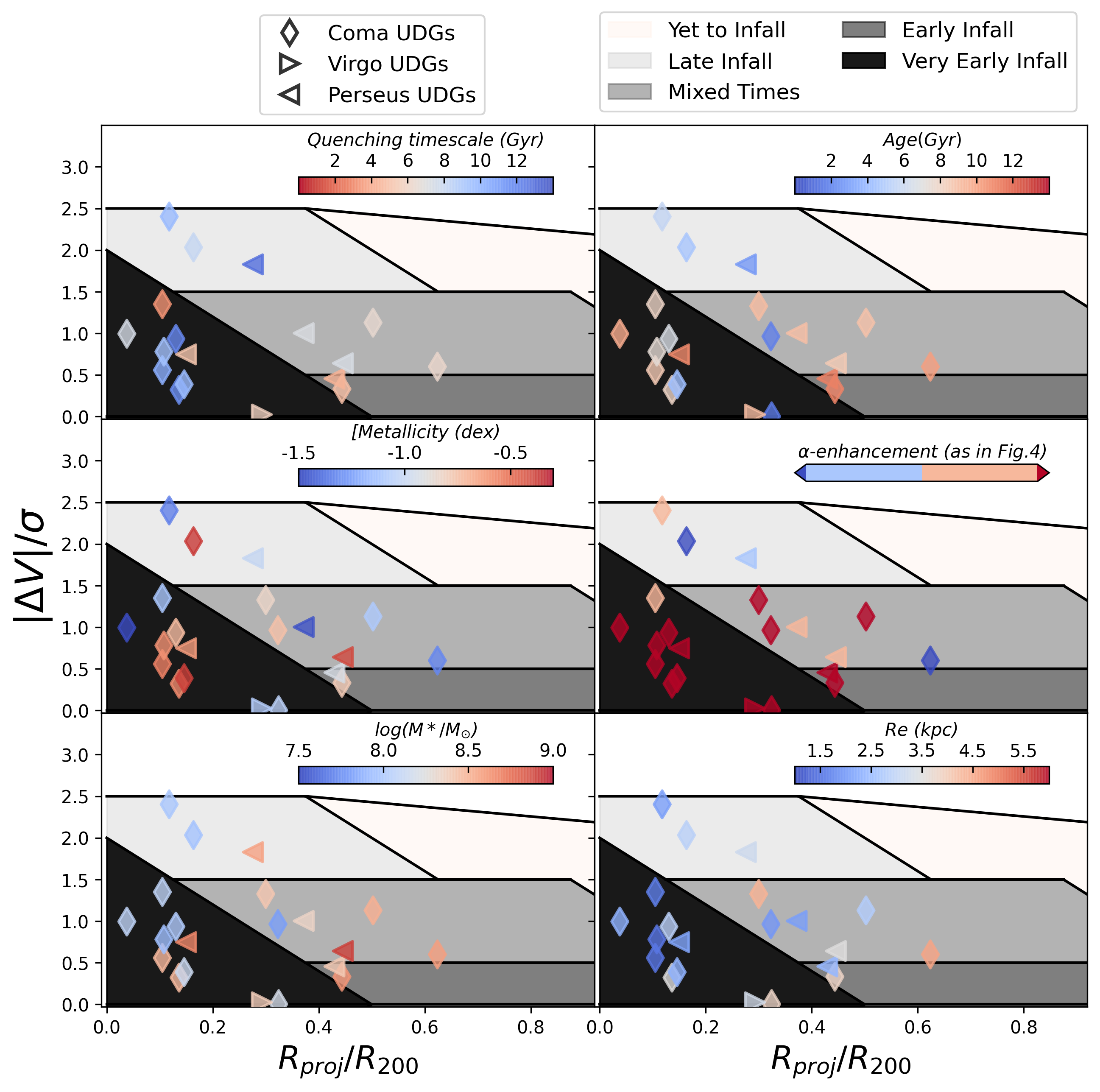}
\caption{Phase-space diagram for the full spectroscopic sample of cluster UDGs. In each panel, the different infall regions from \citet{Rhee2017} are shown as shaded regions, going from darker to lighter to show UDGs infalling at progressively later times. In each panel, cluster UDGs are colour-coded by the respective property (i.e., quenching time, age, [M/H], $\alpha$-enhancement, stellar mass and half-light radius), as shown in each colour scale. We note that quenching time and stellar age have opposite colour scales to guide the eye and that the color scheme in the $\alpha$-enhancement panel follows that presented in Figure \ref{fig:sfhs_all_fig}. Different symbols are used to differentiate the cluster each UDG belongs to. No clear trends are seen between infall regions and most of the UDG properties. We find that very early/early infall UDGs mostly present more elevated $\alpha$-enhancement values than late infalls, compatible with their SFHs.} 
\label{fig:phasespace_fig}
\end{figure*}

\subsection{Phase-space and infall time of UDGs}\label{section:phasespace}
Both the global (cluster, group or field) and local (center, outskirts, filament, etc...) environment appear to influence the evolution of UDGs. For example, in \hyperlink{FM+18}{FM+18} it was found that the stellar populations of UDGs correlate with clustercentric distance. UDGs in the outskirts were younger, most likely related to a later cluster infall time. As a consequence, they had later quenching times and thus presented more extended SFHs (see also \citealt{Alabi2018}). 

In Figure \ref{fig:phasespace_fig} we provide a position--velocity phase-space study of our UDG sample, in order to test whether or not the different stellar populations in the cluster UDGs correspond to their location in a phase-space diagram. It shows the velocity of the galaxy compared to the mean velocity of each cluster (normalised by the velocity dispersion of the cluster) versus the projected clustercentric distance (normalised by the virial radius, R$_{200}$). For each cluster analyzed here we assume: R$_{200}$=2.9\,Mpc, $V_r$=6943~\kms\ and $\sigma$=1031~\kms\ for Coma;  R$_{200}$=1.7\,Mpc, $V_r$=1137~\kms\ and $\sigma$=752~\kms\ for Virgo; and R$_{200}$=1.8\,Mpc, $V_r$=5366~\kms\ and $\sigma$=1324~\kms\ for Perseus (\citealt{Alabi2018}; \citealt{Gannon2022}). Over-plotted are the cluster UDGs from Table \ref{table:bonafide}, with symbols representing the different clusters and the colour scale in each panel corresponding to the relevant properties (when available). 

In Figure \ref{fig:phasespace_fig} there is also overlaid a scheme that correlates phase-space location with the different infall epoch from the simulations of \citet{Rhee2017}. The shaded sections go from a dark to a light shade, corresponding to the transition from ancient to more recent infalls, (i.e., `very early infall', `early infall', `mixed times', `late infalls' and `yet to infall'). We note, however, that these regions are statistically significant only with large samples of galaxies. In all regions there could be large numbers of interlopers: galaxies that do not belong to the cluster but that projection effects place them within projected cluster phase space, or cluster galaxies with a late infall time that are also placed in the center due to projection effects. 

To begin, we wish to confirm the validity of the infall regions, by comparing them to the quenching timescale inferred from the SFHs of the UDGs. Assuming that cluster infall is the cause of quenching, the infall time can be estimated as t$_\mathrm{inf}\sim$t$_{90}-1.5$\,Gyr. However, other UDG formation scenarios invoke self-quenching processes. This means that some UDGs might have formed and quenched before cluster infall, in which case early quenched galaxies in late infall regions can be found. According to \citet{Rhee2017}, ancient infallers are considered to be galaxies that took up to 7.25\,Gyr to fall in the cluster, roughly conforming the `very early infall' and `early infall' regions. 9 UDGs (PUDG-R27, VCC~1287, Yagi358, Yagi093, Yagi098, Yagi276, Yagi392, Yagi418 and OGS1) are located in the `very early infall' region, with other 2 (DF07 and PUDG-R15) in the `early infall' region. They have quenching timescales compatible with their location with the exception of Yagi093, Yagi098, Yagi276 and Yagi418, which present very long quenching timescales that are more similar to those expected for intermediate infallers (timescales of 7.25 to 10\,Gyr). These UDGs could therefore be the result of a projection effect (i.e., interlopers). 4 UDGs are located in the `mixed times' regions (PUDG-S74, PUDG-R84, DFX1, DF44) all presenting quenching timescales of $\sim$6--8\,Gyr. Finally, galaxies in the  `late infall' region should have taken from 10 to 13.7\,Gyr to infall in the cluster according to \citet{Rhee2017}. 3 UDGs (PUDG-R24, Yagi275 and Yagi090) are located in this region, in agreement with their long quenching timescales and relatively young ages. We note that for DF17, J130038.63+272835.3 and J130026.26+272735.2, we do not have an estimate of the quenching time and they are thus not included in this first panel. 
 
Typically, low-mass galaxies that have been in the cluster environment for longer are found to be predominantly old and with higher \afe\, abundance ratios than recently accreted ones (e.g. \citealt{Liu2016}; \citealt{Pasquali2019}; \citealt{Smith2009coma}; \citealt{Gallazzi2021}; \citealt{Bidaran2022}). 
We find that the majority of the UDGs in the `very early infall' and `early infall' regions show indeed elevated $\alpha$-enhancement values (following the scheme presented in Figure \ref{fig:sfhs_all_fig}). Interestingly, these UDGs also present some of the highest metallicities in the sample, similarly to what was found in \citet{Bidaran2022} for classical dwarfs. Conversely, low $\alpha$-enhancement ratios are only found in mixed times towards late infall regions, compatible with the SFHs trends found in the previous section. There are two UDGs whose stellar populations and phase-space do not seem to match. Yagi090, one of the very late infallers, presents a very high [Mg/Fe] \citep{Ruiz-Lara2018}. According to its SFH it was quenched 5\,Gyr ago, which would imply that this UDG was quenched prior to infall. DF44 instead presents a relatively low [Mg/Fe] \citep{Villaume2022} even though it is located in the `early infall' region. However this UDG has been reported to belong to a group that is currently plunging through the Coma Cluster, and thus has been previously considered to be a low-density UDG. We note that this dependence of the $\alpha$-enhancement with galactocentric distance (and lack of thereof for the ages and metallicities) is similar to the one recently reported by \citet{Romero-Gomez2023} for dwarfs in Fornax. No other clear trends between the phase-space infall times and the structural properties of UDGs are observed.

\begin{figure*}
\centering
\includegraphics[scale=0.29]{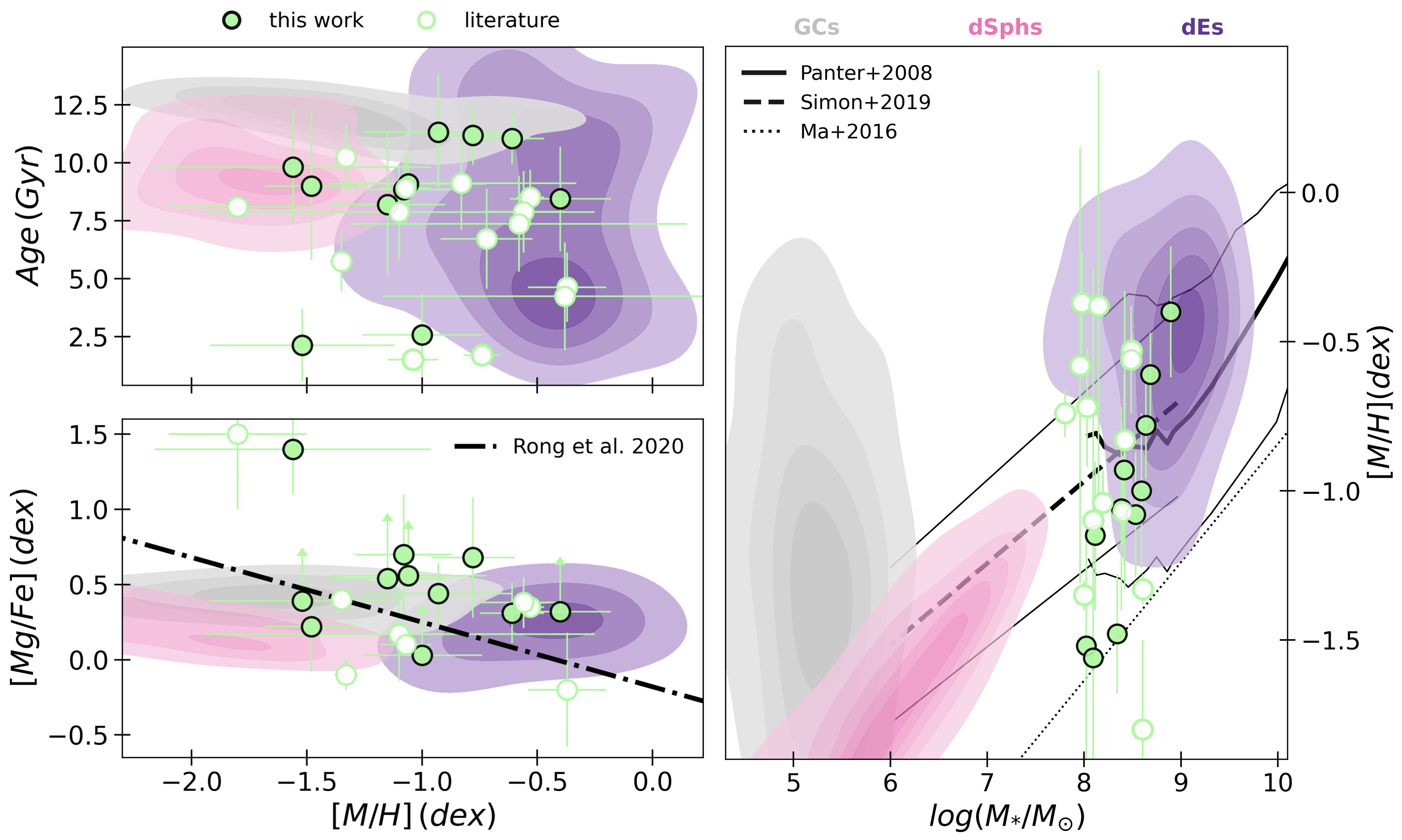}
\vspace{-0.1cm}
\caption{UDG scaling relations compared to other low mass galaxies and GCs. The figure shows the relation of the UDG metallicities with stellar age (top left), [Mg/Fe] (which includes the relation for UDGs defined in \citealt{Rong2020a}; bottom left), and the well-known mass--metallicity relation (right panel). For the latter, the scaling relations and their intrinsic scatter of local massive and low-mass galaxies (\citealt{Panter2008}; \citealt{Simon2019}, respectively) and the theoretical prediction at $z\sim 2$ \citep{Ma2016} are also shown. UDGs from this work are plotted in filled symbols, and UDGs from the literature with open ones. Contours with different colours correspond to the locations of samples of GCs (grey), Local Group dSphs (pink) and Virgo/Fornax/Coma dEs (purple), from \citet{Janz2016}; FM+18; \citet{Recio-Blanco2018}; \citet{Naidu2022}; and \citet{Romero-Gomez2023}. We have marked the UDGs with the $\alpha$-enhancement obtained from \ppxf as possible lower limits by arrows. UDGs show overall elevated [Mg/Fe], in some cases much higher than any other galaxy type known. UDGs show a large dispersion in their stellar ages and metallicities, scattering around the local mass--metallicity relation. A handfull of UDGs seem to be more compatible with high-$z$ galaxies, suggesting these could have a different origin (i.e. `failed-galaxy' like).}
\label{fig:rels_allgals_fig}
\end{figure*}

\subsection{Stellar populations trends}\label{section:stellarpopstrends}
We next try to elucidate the most plausible origin of the spectroscopic sample of UDGs by comparing them to other low mass systems that have been proposed to be related. For example, GCs are broadly understood to be old and metal-poor, having formed during the early stages of galaxy formation. Some works suggest that UDGs form in the same epoch as the GCs (i.e., the failed-galaxy type; \citealt{Forbes2020c}) or that they may even form entirely from disrupted clusters (i.e., NGC~5846-UDG1; \citealt{Danieli2022}), so any UDG that shares similar properties as GCs could be considered of this type. On the contrary, if UDGs are simply puffed up dwarf galaxies, one would expect to find very similar stellar populations to classical dwarfs but with possibly different stellar masses. 

Figure \ref{fig:rels_allgals_fig} presents the age--metallicity relation, the [Mg/Fe]--metallicity relation, and the mass--metallicity relation, some of the most relevant scaling relations regarding the stellar populations of galaxies. In the [Mg/Fe]--metallicity panel the relation that was derived for UDGs is shown \citep{Rong2020a}. We include it as it was obtained using mostly quiescent UDGs, although star forming isolated UDGs were also included in the fit.  In the mass--metallicity panel the scaling relation of \citet{Panter2008} is shown, which is an extension of the \citet{Gallazzi2005} relation of massive ETGs towards intermediate stellar masses (roughly covering the UDG mass regime). The theoretical mass--metallicity relation of high redshift galaxies ($z\sim2$) from \citet{Ma2016} is also shown. 
The reader is referred to \citet{Buzzo2022} for an extensive discussion on the caveats of the different mass--metallicity scaling relations at the low mass end, here we use the relation from \citet{Simon2019}. In each panel, the coloured contours correspond to different types of low-mass systems: GCs, dSphs and dEs (spectroscopic values from \citealt{Janz2016}; \hyperlink{FM+18}{FM+18}; \citealt{Recio-Blanco2018}; \citealt{Naidu2022} and \citealt{Romero-Gomez2023}). These control galaxies cover a range in environments: from the Local Group to clusters (Fornax, Virgo, Coma). dSphs have lower stellar masses than the UDGs, and the dEs have mostly higher masses.

The age--metallicity panel of Figure \ref{fig:rels_allgals_fig} (top left) shows that the majority of UDGs have ages and metallicities similar to dEs for the more metal-rich UDGs, while more metal-poor ones share similar age--metallicities to dSphs. Three UDGs (PUDG-R15, PUDG-R27 and DF07) match the GCs distribution, although in a region that also matches the dEs. A few other UDGs would be compatible with the GC distribution given the errors in metallicity. We note, moreover, that some GCs in the control sample have ages of $\sim$7-8Gyr and [M/H]$\sim$-0.8\,dex, similar to dEs. However, they are so few compared to the bulk of GCs that are not shown by the contours. UDG1137+16 is the only UDG that is clearly different to any of the comparison samples, with a metallicity that is more typical for older galaxies. This UDG is in fact an outlier in the majority of the scaling relations, which might be a consequence of the tidal features that it shows (see e.g. \citealt{Gannon2021}).

In the [Mg/Fe]--metallicity panel (bottom left) it is shown that most UDGs are also very $\alpha$-enhanced, some of them more than any other known galaxy of the comparison distributions. Only five UDGs in the sample present scaled solar or even sub-solar [Mg/Fe]. In this panel we also include the [Mg/Fe]--metallicity relation derived for UDGs \citep{Rong2020a}. However, we find that the bulk of our UDGs is more compatible with a flat relation, similar to what is seen for the comparison samples. We remind the reader that this is mainly caused by the limitations discussed in Section \ref{section:alphas} and the Appendix \ref{section:ap_alfas} on the models and methods used to obtain this quantity. UDGs mostly cover the [Mg/Fe] range of GCs and dwarfs at all metallicities, and only the most extreme ones (DGSAT~I and Yagi358) having [Mg/Fe] that do not match any galaxy in the comparison sets. In particular, DGSAT~I has been shown to be an extremely bizarre galaxy in many respects \citep{Martin-Navarro2019}. Its high [Mg/Fe] value has been potentially explained by a `depletion' of [Fe/H] rather than an over-abundance of [Mg/H]. From our line index analysis we also measure an extremely low [Fe/H] but high [Mg/Fe] for Yagi358. Given that they both have overall similar stellar population properties, we speculate that Yagi358 could have formed in a similar process to DGSAT~I.

The right panel of Figure \ref{fig:rels_allgals_fig} shows different mass--metallicity relations. Overall, UDGs in this sample scatter around the mass--metallicity relation expected for galaxies of log(\Mstar/\msun)$\sim$8--9. Despite the large scatter in the metallicity distribution, UDGs seem to have on average lower metallicities than the dEs at fixed stellar mass. 
Five UDGs present much lower metallicities than the rest of UDGs and classical dwarfs, laying below the local mass--metallicity relations. These are Yagi358, PUDG-R84, DF44 and DGSAT~I. DGSAT~I has been already reported to be an unusual galaxy, being an extreme outlier of most scaling relations (e.g. \citealt{Martin-Navarro2019}). As the failed-galaxy formation scenario assumes early quenching, UDGs formed this way are expected to follow the mass--metallicity relation of high-$z$ galaxies rather than local ones. We find that these other four UDGs with very low metallicities indeed match the theoretical mass--metallicity relation at z$\sim$2 \citep{Ma2016}, which would support them as failed-galaxies. 

\subsubsection{Trends with environment and GC-richness}
We have seen in Section \ref{section:sfhs} a dependence between the local environment and the stellar populations of UDGs and we have shown in Figure \ref{fig:rels_allgals_fig} that the bulk of UDGs in this sample have mostly stellar population properties similar to classical dwarf galaxies. While this is compatible with a puffy dwarf origin, it is difficult to explain their populous GC systems, as reported for many cases. For example, \citet{Lim2018} and \citet{Forbes2020c} showed that Coma UDGs tend to have higher GC specific frequencies than comparable classical dwarf galaxies. Additionally, it is not clear whether there are any differences in the overall properties of GC-rich and poor UDGs. For instance, a trend with GC-richness was found by \citet{Buzzo2022}, with GC-poor UDGs presenting higher metallicities than GC rich ones. From their SED fitting they also found that the oldest UDGs tend to prefer cluster environments, supported by our results in Figure \ref{fig:sfh_byprop}.

\begin{figure*}
\centering
\includegraphics[scale=0.28]{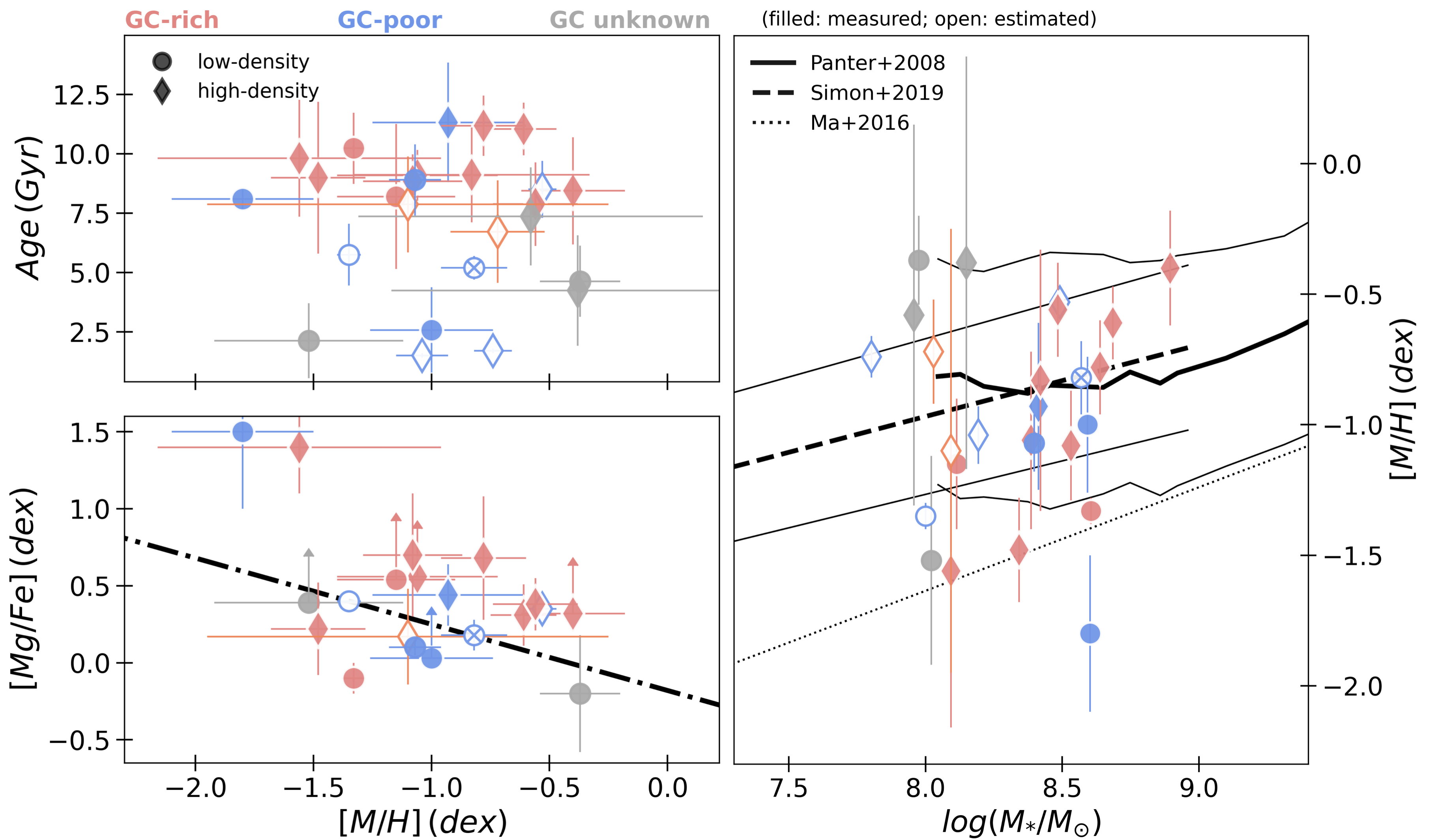}
\caption{Main scaling relations for the total sample of UDGs with GC-richness and environment. Same as in Figure \ref{fig:rels_allgals_fig} but now UDGs are color coded by GC richness: red for GC-rich, blue for GC-poor, and grey for UDGs with no GC information. Filled symbols correspond to UDGs with proper GC counts, open symbols if the GC-richness comes from the visual estimate described in Appendix \ref{section:ap_gcrich}. Circles correspond to low-density UDGs and diamonds to high-density ones, classified as in Figure \ref{fig:sfh_byprop}. The stacked value for the star forming, isolated UDGs of \citet{Rong2020a} is marked with an X in its symbol. We find that GC-rich UDGs have intermediate-to-old ages, while the young ones are all GC-poor. However we do not find any evident trends of GC richness with [Mg/Fe] nor in the mass--metallicity relation. }
\label{fig:rels_richenv_fig}
\end{figure*}

Therefore we investigate next if similar trends are recovered in this spectroscopic sample. Figure \ref{fig:rels_richenv_fig} shows the sample of UDGs now separated between low density (field and group UDGs, but also UDGs considered to be recent infalls), and high density environments (the rest of cluster UDGs), as it was done in Figure \ref{fig:sfh_byprop}. To complement our comparisons within environments, we also include in this figure the stellar populations properties measured for a sample of stacked isolated, star forming UDGs (crossed circle, \citealt{Rong2020a}), as one formation scenario for UDGs predicts that these may evolve from star-forming in the field to quenched in the cluster environment. The color scheme describes the GC-richness of the UDGs, where we use both proper counts but also UDGs with a visual classification (see Appendix \ref{section:ap_gcrich}). 

We find that overall, most (13/16) high density UDGs show intermediate-to-old ages ($\gtrsim$7\,Gyr), confirming the results from Figure \ref{fig:sfh_byprop}, now with an expanded sample. Young UDGs ($\lesssim$ 7 Gyr) are found evenly in terms of local environment. This result is thus further compatible with \citet{Buzzo2022}, although their sample did not include UDGs with such young ages. 

Similarly to \citet{Buzzo2022}, we find that the UDGs lying on the mass--metallicity relation of high-$z$ galaxies are all GC-rich (without considering DGSAT-I, which has been repeatedly reported as an unusual galaxy). Having a large number of GCs is one of the main expectations of the failed-like galaxy scenario, further supporting such origin for these UDGs. However, we do not find any trend between GC richness and galaxy metallicity as reported in \citet{Buzzo2022}. In the latter, GC-poor UDGs have higher metallicities than GC-rich ones, while we have many GC-rich UDGs with elevated metallicities in this study.

There are many possibilities to explain these differences. For example, there seems to be a systematic effect towards lower metallicities in the SED fitting compared to spectroscopy (see Appendix \ref{section:ap_complit}). It is unclear if the impact would be the same for UDGs at all ages, but we find that in particular the UDGs with the largest difference in metallicity are the oldest ones (Figure \ref{fig:sedVSspe_fig}). There could also be a sample bias, as there are no GC-rich UDGs that are metal-rich in the work of \citet{Buzzo2022}, while in this work there are several. We note that in \citet{Barbosa2020}, also from SED fitting but targeting field UDGs, a large number of UDGs with such higher metallicities were reported, but unfortunately no GC counts are yet available. We also note that our sample is biased towards both high-density environments and GC-rich UDGs. In addition, most of the GC-poor ones are coming from a visual analysis, and not from proper GC counts. Nonetheless, we chose to use the visual GC counts as we are primarily interested in a comparative study of the properties of UDGs with/without GCs, not on their absolute numbers. We note, however, that the trend is still not found even if we exclude these galaxies from the analysis, hence further work studying the GC systems of these galaxies is necessary to understand if our findings hold.

Additionally, no particular trend is seen with the $\alpha$-enhancement of the UDGs and their GC system richness. At most, it seems that GC-poor UDGs follow the \citet{Rong2020b} relation more tightly than GC-rich ones (with the exception of DGSAT-I, again). Only two out the six low-density UDGs with a GC estimate lie above the [Mg/Fe]--metallicity relation, an area mostly populated by the high-density UDGs.

Bearing in mind the aforementioned caveats, the take away message is that until we have larger and more complete samples of UDGs of varying GC richness and environments, any trends obtained may be simply an effect of the sample selection itself and must be taken with caution for now.

\begin{figure*}
\centering
\includegraphics[scale=0.85]{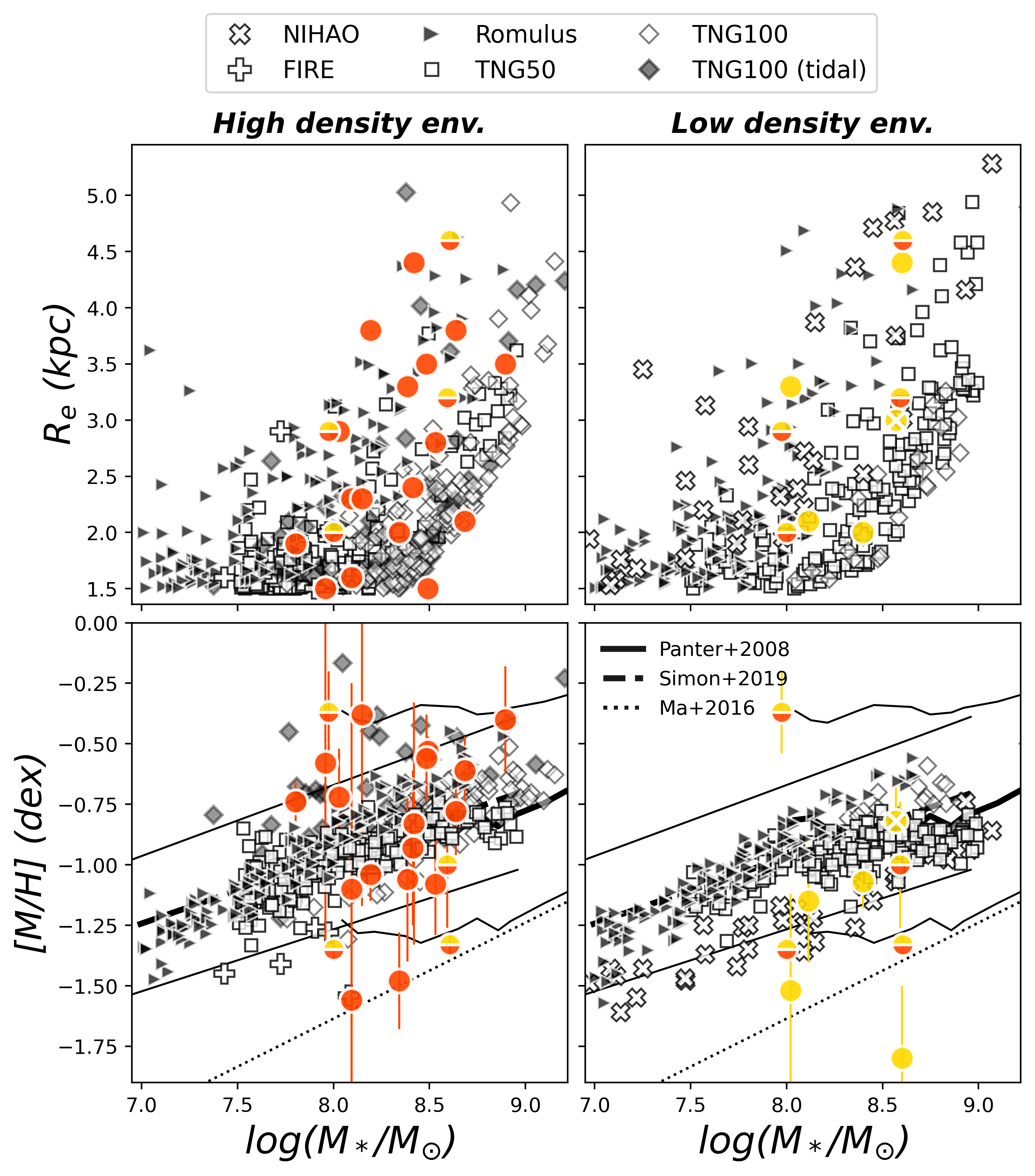}
\caption{Comparison of observed and simulated UDGs separated by environment: high-density (cluster, left column) and low-density (field, group; right column).
The following sets of simulations cover high density environments:  FIRE (white crosses; \citealt{Chan2017}), Romulus (black triangles; \citealt{Tremmel2020}), TNG50 (white squares; Benavides private communication) and TNG100 (white diamonds for cluster UDGs, filled in grey for those that are considered to be tidal-stripped galaxies from \citealt{Sales2020}). Romulus, TNG50 and TNG100 are also shown in the low-density section, together with the UDGs from NIHAO (white x-symbols; \citealt{DiCintio2017}). The top panels present the mass--size relations, where simulated UDGs are limited to \re$\ge$ 1.5\,kpc to match the observations, and TNG50 metallicities have been re-scaled as in \citet{Sales2020}. Local scaling relations (solid lines) at the high mass \citep{Panter2008} and low-mass ends \citep{Simon2019}, and their scatter (thin solid line), are shown. The high redshift theoretical mass--metallicity relation of \citet{Ma2016} is also shown in the bottom panels (dotted line). The spectroscopic sample of UDGs is represented by circles, coloured by their local environment: orange for UDGs in high density environments, yellow for low density ones. Circles with half-and-half colours denote UDGs thought to be very recent cluster infalls or located in filaments. These are shown in all panels, as their properties could be similar to both field UDGs (before infall) and cluster ones. Like in the previous figure, the properties of the isolated, star forming UDGs from \citet{Rong2020b} are marked by a white X. It is clear that there is not a single simulation set that reproduces all type of UDGs at the same time. UDGs with the lowest metallicities and that lay below the local mass--metallicity relation (i.e., most likely failed-galaxy type), are the only UDGs not reproduced by any set of simulations.}
\label{fig:simul_fig}
\end{figure*}

\subsection{Comparisons to simulated UDGs}\label{section:simuls}
Many different simulations have tried to reproduce the observed properties of UDGs by proposing a myriad of formation mechanisms. Here we compare some of the observed properties discussed in this work to different sets of simulated UDGs. With this exercise we wish to have a qualitative assessment of the compatibility of simulations and observations. We note, however, that most of the simulations that form UDGs do {\it not} form the more compact classical dE/S0s, making up the vast majority of observed dwarfs in the same mass range -- an example of the `dwarf diversity' challenge to current models for galaxy formation \citep{Sales2022}.

The first simulation we compare to is the NIHAO cosmological simulation \citep{Wang2015}. The UDGs from this simulation have been studied in detail by \citet{DiCintio2017}, \citet{Jiang2019} and \citet{CardonaBarrero2020}. These simulations create UDGs in low-desnity environments through strong stellar feedback that puffs up dwarf galaxies. The simulated UDGs present bursty SFHs, intermediate ages and low metallicities (\citealt{DiCintio2017}; \citealt{CardonaBarrero2020}). 

We also use UDGs from the FIRE simulations of \citet{Chan2017}. In this case, the simulation mimics the creation of UDGs in a cluster environment by quenching the simulated UDG at a given time. Two different scenarios were investigated: one where the galaxy is artificially quenched very early on (roughly $\sim$2~Gyr after the Big Bang), while the other scenario allows UDGs to evolve for longer times, quenching only about 4\,Gyr ago to simulate late cluster infalls. Similar to NIHAO, FIRE primarily creates UDGs through stellar feedback before their artificial quenching. 

We include simulated UDGs from the Romulus cosmological simulation \citep{Tremmel2017}, which are all considered to be like puffed-up dwarfs. The sample consists of a set of cluster UDGs (from RomulusC, \citealt{Tremmel2020}) and isolated UDGs (from Romulus25, \citealt{Tremmel2017}, \citealt{Wright2021}). Those in clusters present different infall times, although the authors found that UDGs seem to fall into the cluster earlier than regular dwarfs. Those in the field are the result of major mergers occurring at earlier stages than non-UDGs, increasing their spin and redistributing the star formation towards the outskirts. Therefore for both cluster and field UDGs in the Romulus simulations, their nature is finally shaped by the passive evolution of their stellar populations \citep{Roman2017a}. 

We also include two different sets of Illustris-TNG simulations. We employ the simulated UDGs of Illustris-TNG50  \citep{Benavides20222}, which are a mixture of high and low density environments (sampling cluster, groups and field UDGs). These also include the backsplash UDGs presented in \citet{Benavides2021}. Overall, these authors found that cluster UDGs tend to be smaller and less massive than field UDGs or recent infallers, with slightly lower metallicities. We also include the sample of \citet{Sales2020}, which presented UDGs from the Illustris-TNG100 simulation. These are classified as either `born' UDGs (already an UDG before cluster infall), or those that `become' a UDG due to tidal effects within the cluster after infall (`tidal UDGs'). The latter are typically accreted earlier in the cluster (in 1 to 6 Gyr after the Big Bang) and have larger stellar masses at the time of the infall. We caution the reader that all TNG simulations have some difficulties in reproducing the metallicities of galaxies (e.g., \citealt{Nelson2018}; \citealt{Sales2020}; \citealt{Benavides2021}). For instance, TNG50 simulations were originally re-scaled to match the observed scaling relations at log(\Mstar/\msun)=9 (see \citealt{Benavides20222}). The shift applied ($\sim$0.5\,dex) is comparable to the reported discrepancy in \citet{Nelson2018} for all galaxies in TNG100. Therefore, we apply the same re-scaling to the TNG100 sample of UDGs (Benavides et al. priv. comm.).

Figure \ref{fig:simul_fig} shows the stellar mass--size relation (top) and the mass--metallicity relation (bottom), two of the main scaling relations studied in this work. We plot the corresponding simulated and observed UDGs according to their local environment: high-density environments (clusters; left panels) and low-density ones (groups and field; right panels). UDGs that are located in the `late-infall' region of Figure \ref{fig:phasespace_fig} or that are known to be in an infalling group, such as DF44, are shown in all panels. The mass--size panels are simply shown to have all set of simulations together with the observed UDGs.  For all sets of simulations we have only considered UDGs that match the observational criteria of \re$\ge$1.5\,kpc. Many of the simulations produce UDGs with lower stellar masses than in our observed sample, in particular in low-density environments. This is most likely the result of spectroscopic observations being biased towards more massive UDGs. 

In the stellar mass--metallicity relations of Figure \ref{fig:simul_fig}, most simulations predict a relation that changes only very weakly with environment -- similar to observations of a near-universal red sequence for classical dwarfs. Most of the observed UDGs in high-density environments share location with the simulated UDGs (lower left panel), while UDGs in low-density environments, as well as late infallers, are on average offset to lower metallicities (lower right). 

NIHAO has a mass -- metallicity relationship in their simulations systematically lower than that of other simulations and what is observed in this stellar mass regime. It is not clear that NIHAO can produce non-UDGs following the mass-metallicity relationship in this mass regime. This is a result of their strong internal feedback, which expands the galaxies and can expel the metals. Despite the systematically lower metallicities, NIHAO does not produce UDGs as low as some of the observed UDGs (e.g., that match the theoretically z$\sim$2 mass--metallicity relation). Furthermore, this stellar feedback mechanism was rejected to explain the formation of the star forming isolated UDGs in \citet{Rong2020b} as these match the expected scaling relations and are therefore better represented by other simulations.

It is important to note that there is not one single formation scenario proposed for UDGs and that there is evidence of UDGs forming via alternate pathways in Figure \ref{fig:simul_fig}. For example, Yagi275 and Yagi276 are two of the UDGs with the highest metallicities for their stellar mass. One possibility to explain galaxies above the relation is that such elevated metallicities are indicative of a tidally stripped galaxy (e.g. \citealt{Gallazzi2005}; \citealt{Ferre-Mateu2018b}; \citealt{Du2019}; \citealt{Ferre-Mateu2021}). Interestingly, these two UDGs are located where the simulated UDGs resulting from tidal effects are found (in the high-density environment panel; \citealt{Sales2020}). This scenario would be plausible for Yagi276, as it has long resided in the cluster (owing to its infall diagram), which is one of the requirements of this scenario. However, it would only work for Yagi275 if this UDG was already a stripped galaxy before infall, as it is considered to be a late infall from both its phase-space location and SFH.

The most interesting cases are the UDGs that present the lowest metallicities, laying below the local mass--metallicity relation but follow instead the theoretical relation expected for high-$z$ galaxies. From those, PUDG-R84 and Yagi358 (in high-density environments), and DF44 (as infall galaxy), present early and fast formation and are GC-rich,  indicative of a large dark matter halo. These three UDGs are therefore the best candidates to failed-galaxy like UDGs considering all their inferred properties (see also previous sections). Interestingly, these three UDGs are not reproduced by any simulation in the high-density environment. The only simulated galaxy compatible with this scenario is one of the FIRE UDGs that was forced to quench very early on \citep{Chan2017}. This further supports the failed-galaxy origin for these subset of UDGs. 

DGSAT-I and UDG1137+16 (both in low-density environments), although they also present low metallicities and lay below the relations, do not met the rest of expectations for being considered failed-galaxy like UDGs. The uniqueness of the exotic chemistry of DGSAT-I has already been discussed (e.g., \citealt{Martin-Navarro2019}) but not reproduced yet in simulations. UDG1137+16 has a complicated formation as shown by its tidal signatures \citep{Gannon2021}, but it has metallicity similar to the NIHAO UDGs of similar stellar mass. 

We have at hand several simulations that attempt to create UDGs following different prescriptions, although there is not a single simulation that can reproduce the diversity of observed UDGs. In addition, many present caveats (e.g. can not reproduce metallicities and stellar masses of observed UDGs). Finally, galaxies with properties similar to UDGs of failed-galaxy origin have not successfully been simulated as of yet.  

\section{Summary} \label{section:summary}
In this work we have analyzed the stellar populations of the largest spectroscopic sample of observed UDGs to date (25), covering different environments and GC richness. This sample comprises 11 UDGs for which new spectroscopy from KCWI on the Keck Telescope was obtained, along with 14 literature UDGs. The new sample is mostly comprised of UDGs in Perseus, Coma, Virgo and two UDGs in groups of galaxies. For all UDGs in our sample, we obtain their structural properties and number of GCs from published works (\citealt{Gannon2020G}; \citealt{Forbes2021}; \citealt{Gannon2021}; \citealt{Gannon2022} and \citealt{Gannon2022b}). For some of the UDGs without previous GC counts, we have performed a visual inspection to increase the sample, although we caution the reader about using these estimates at face-value. We have carried a new stellar population analysis for the UDGs and discussed different relations and dependencies in order to elucidate their most plausible origin (see Table \ref{table:origins} for a summary of all properties and suggested origins). 
\begin{itemize}
    \item We have derived the main stellar population properties (such as mean ages, metallicities and $\alpha$-enhancement) of these 11 new UDGs via both a full spectral fitting approach (using \ppxf) and using classical line index measurements. These UDGs have a mean age $\langle \,t_\mathrm{M}\,\rangle$=8.3$\pm$3.3\,Gyr, metallicity $\langle$[M/H]$\rangle=-1.03\pm0.37$\,dex, and mean $\alpha$-enhancement of $\langle$[Mg/Fe]$\rangle=+0.51\pm 0.32$\,dex. 
    \item We find a clear trend between the SFH and the $\alpha$-enhancement. UDGs that formed earlier and faster present the highest enhancements, while those forming over a longer period of time show scaled solar values. UDGs with intermediate ages also show relatively high abundance patterns because they formed relatively quick despite beginning their star formation later in time. 
    \item Local environment seems to be relevant in the way UDGs are formed. UDGs in high density environments show very early onset of star formation, quenching within timescales of $\sim$6\,Gyr. All these UDGs are at the high mass end of the UDG population. On the contrary, UDGs in low density environments (groups and cluster outskirts), started forming their stars roughly 10\,Gyr ago, and did not quench until very recently.
    \item We thus propose different formation scenarios for the UDGs according to their environment and stellar populations. UDGs that formed early and quickly, thus quenching at early ages, can be considered ancient infalls into the cluster. These can be UDGs with a failed-galaxy origin if their metallicities are low, or tidally stripped UDGs if their metallicities are high. The rest of UDGs in the sample are more compatible with having a puffy dwarf origin. 
    \item We study the phase-space for cluster UDGs related to the main stellar population properties such as age, metallicity, $\alpha$-enhancement and estimated quenching times, while looking for trends with the stellar mass and size of the UDGs. We find that the location in the phase-space diagram only hints for a relation with [Mg/Fe] but not with any other property. We therefore caution about using such phase-space diagrams to determine the true locations in the cluster. 
    \item Many UDGs share similar properties (ages and metallicities) to dEs, suggesting a puffy dwarf-like origin for them. However, a subset of UDGs could have been formed as failed galaxies, as they present the lowest metallicities and lay below the local scaling relations but follow the high-$z$ ones instead.  
    \item We include the information of the GC richness to investigate previous trends reported. We find that all GC-rich UDGs in our sample have intermediate-to-old ages and that all the young UDGs are GC-poor. Moreover, we find that the UDGs suggested as failed-galaxy UDGs are all GC-rich, further supporting this formation scenario for such UDGs. 
    \item UDGs have a large number of simulations trying to reproduce their properties. These simulations are successful in producing the relevant stellar population properties for a large fraction of observed UDGs in this work. However, there is no single simulation that can recreate the observed diversity of UDG properties and many of the simulations are unable to correctly reproduce some of the relevant properties for the non-UDG population. Moreover, the observed UDGs that match expectations of $z\sim$2 galaxies are not reproduced by any simulation as of yet. 
    \end{itemize}

We thus find that there exists a wide range of formation pathways to create a UDG. Failed-galaxies UDGs are one of these pathways, although they appear more rare than those following a puffy-dwarf formation scenario in our data. In order to seek conclusive trends that can help differentiate the various formation pathways of UDGs, we are in need of new, more complete spectroscopic studies targeting UDGs (e.g., GC-poor UDGs, low-density environment UDGs).

\section*{Acknowledgements}
We thank J. Benavides, S. Cardona-Barrero, A. di Cintio, L. Sales, and M. Tremmel for providing the data of the different simulated UDGs, as well as providing insightful discussions. We also thank T. Ruiz-Lara for providing the SFHs for the Coma UDGs published in \citet{Ruiz-Lara2018} and Jorge Romero for the dwarf data from \citet{Romero-Gomez2023}. We also thank the referee for their reports that helped improve the quality of the manuscript.\\

AFM has received support from CEX2019-000920-S, RYC2021-031099-I and PID2021-123313NA-I00 of MICIN/AEI/10.13039/501100011033/FEDER,UE, NextGenerationEU/PRT, and through the Postdoctoral Junior Leader Fellowship Program from `La Caixa' Banking Foundation (LCF/BQ/LI18/11630007). This research was also supported by the Australian Research Council Centre of Excellence for All Sky Astrophysics in 3 Dimensions (ASTRO 3D), through project number CE170100013. AJR was supported by National Science Foundation grant AST-2308390. Support for Program number HST-GO-15235 was provided through a grant from the STScI under NASA contract NAS5-26555. This work was supported by a NASA Keck PI Data Award, administered by the NASA Exoplanet Science Institute. \\

Data presented herein are part of the AGATE collaboration (\textit{Analysis of Galaxies At The Extremes}) and were obtained at the W. M. Keck Observatory from telescope time allocated to the National Aeronautics and Space Administration through the agency's scientific partnership with the California Institute of Technology and the University of California. The Observatory was made possible by the generous financial support of the W. M. Keck Foundation. The authors wish to recognize and acknowledge the very significant cultural role and reverence that the summit of Maunakea has always had within the indigenous Hawaiian community. We are most fortunate to have the opportunity to conduct observations from this mountain.

\section*{Data availability}
The KCWI data presented are available via the Keck Observatory Archive\href{https://www2.keck.hawaii.edu/koa/ public/koa.php}, 18 months after observations are taken.

\bibliographystyle{mnras}
\bibliography{UDG_refs}

\appendix
\section{New globular cluster richness estimates}\label{section:ap_gcrich}
In this paper we collect published GC counts and classify the UDGs between GC-poor and GC-rich following previous works (e.g., \citealt{Gannon2022}). This broad and \textit{arbitrary} limit, which is set at $\sim$20 GCs, aims at separating UDGs with dwarf-like dark matter halos from UDGs with more massive ones \citep{Burkert2020}. We note that for the UDGs with already published values, none of them would be classified differently if accounting for the errors in the number count or if considering the different values published.

However, measuring the number of GCs in a galaxy ideally requires deep, multi-band imaging with excellent image quality. Both {\it HST} and ground-based imaging are used for distances out to $\sim$~20 Mpc, while greater distances normally require {\it HST}. Even so, the brighter GCs around a galaxy can be detected out to $\sim$~100~Mpc given particularly high-quality ground-based imaging, which can be enough for a crude assessment of how populous a GC system is. Firm, quantitative measurements require careful photometry, completeness tests, and detailed modeling of the GC spatial distribution and contamination fraction (e.g.\ \citealt{Janssens2022}).  

Nonetheless, in many cases it is possible to obtain a quick visual evaluation of GC richness by assessing the excess number of point-like sources associated with a low surface brightness galaxy. This approach was taken by \citet{Gannon2022} for UDGs in the Perseus cluster using $gri$ imaging from Subaru / Hyper Suprime-Cam, and validated from detailed analysis of a subset of the galaxies using {\it HST}. We obtained {\it HST}/ACS images in either the F475W or F606W filter from the Hubble Legacy Archive. For training purposes, we used three Coma UDGs (DF02, Yagi358, Yagi275) with {\it HST}-based GC numbers in \citet{Forbes2020}. The visual inspection and votes have been carried independently by several co-authors, reaching a consensus for Yagi098, Yagi392 and Yagi090. Yagi418, OGS1, J130026.26+272735.2 and J130038.63+272835.3 are not as clear and are thus placed within parenthesis in Table \ref{table:bonafide}. We therefore caution the reader about using these estimates until better imaging and a proper process is carried out. We note, however, that the results presented and discussed in this paper stand if we do not include the visual classification.

\section{Stellar population measurements}
In this appendix we discuss the main systematics related to the measurement of the most critical properties in this work, namely mean ages, metallicities and $\alpha$-enhancement ratios. 

\subsection{$\alpha$-enhancement}\label{section:ap_alfas}
\begin{figure}
\centering
\includegraphics[scale=0.63]{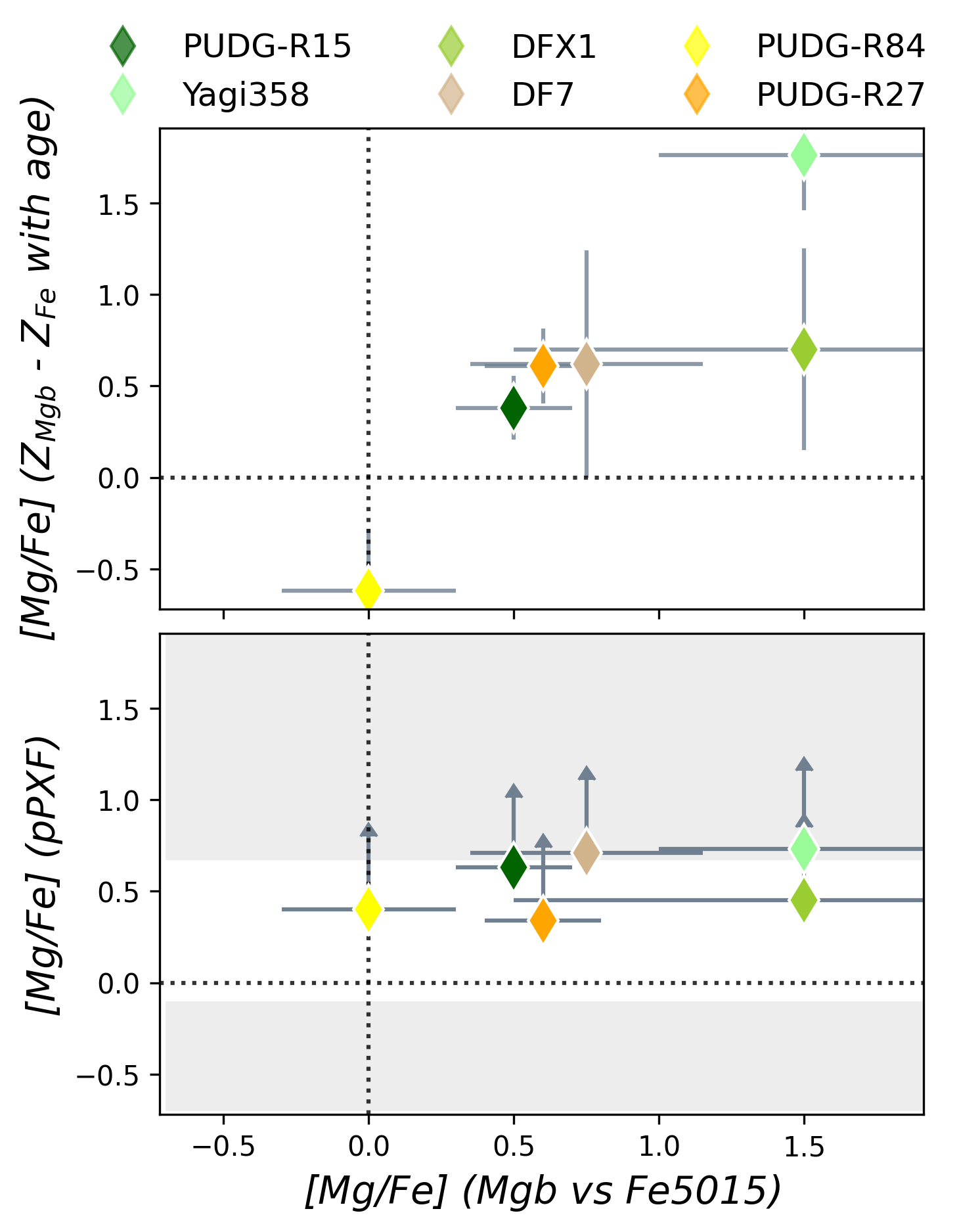}
\vspace{-0.4cm}
\caption{$\alpha$-enhancement measured by different methods for the UDGs with no pedestal issue. \textit{Top}: comparison of the [Mg/Fe] derived from method (i) vs. method (ii): the Mg$_b$ vs. Fe5015 indices for scaled solar and \afe=$+$0.4\,dex SSPs as opposed to age-SSP grids of Mg$_b$ and Fe5015 to obtain the proxy [Z$_{Mg_{b}}$/Z$_{Fe}$]. \textit{Bottom}: comparison of method (i) with the values obtained from the \ppxf routine after converting into [Mg/Fe] (method iii). This last method has the limitation on the SSP models, as shown by the shaded grey areas. The arrows show how much the values could move if a shift is applied, see text.}
\label{fig:alphas_fig}
\end{figure}

We have estimated the $\alpha$-enhancement with three different approaches, as done in other works for faint galaxies with moderate quality spectra (e.g., \hyperlink{FM+18}{FM+18}, \citealt{Ferre-Mateu2021}; \citealt{Forbes2022}). The first method uses the full spectral fitting method (applied to all galaxies), while the other two employ an absorption line index analysis (only for those spectra with no pedestal issue, as discussed in Section \ref{section:alphas}). We therefore obtain three independent estimates:

\begin{enumerate}
    \item using the SSP model predictions with \afe=0.0 and \afe=$+$0.4\,dex  we obtain the [Mg/Fe] from the comparison of the metal indices Mg$_b$ $vs.$ Fe5015. While it is easy to interpolate within the two model grids, extrapolations outside of it make it more difficult. We therefore set as limits for a safe extrapolation of \afe=[$-$0.1,$+$0.6] dex.
    \item using two metallic indices, Mg$_b$ and Fe5015 separately and fixing the ages of the galaxies to those obtained from \ppxf. This provides a more orthogonal grid that does not typically need any extrapolation. From this we obtain the relative metallicities, Z$_{Mg_{b}}$ and Z$_{Fe}$, which are used to obtain the proxy [Mg/Fe]=Z$_{Mg_{b}}-$Z$_{Fe}$. 
    \item using the full spectral fitting approach, we run \ppxf with a set of SSP models that range from scaled solar to \afe=$+$0.4\,dex. This method will always deliver values only within the SSP models used. We note that because this is applied to the entire spectra, in principle will contain several $\alpha$-elements. Therefore this value is closer to a 'true' \afe\, and can be converted into [Mg/Fe] using the formula from \citet{Vazdekis2015} \afe=0.02+0.56$\times$[Mg/Fe].
\end{enumerate}

Figure \ref{fig:alphas_fig} shows a comparison of the different [Mg/Fe] obtained from the three methods. It shows the results from method (i) compared to method (ii) in the top panel, and from method (i) compared to method (iii) in the bottom one. The grey regions in this panel show the parameter space that can not be obtained in the full spectral fitting method. We have marked as in Figure \ref{fig:rels_allgals_fig} the lower limits for the \ppxf estimates, if we were to apply a shift to compensate for this method. This shift has been obtained as the mean difference between the two estimates, and it is roughly +0.25\,dex. Although the results from the line indices are compatible at a first order, all this highlights the need to obtain the [Mg/Fe] abundance ratios from a line index analysis rather than the using full spectral fitting, when possible. For the work presented here we will use the values that are obtained from method (i) for those UDGs with no pedestal issue and the results from \ppxf for the remaining (marked in Table \ref{table:ssp}). 

\subsection{Mean ages and metallicities}\label{section:ap_agemet}
\begin{figure}
\centering
\includegraphics[scale=0.45]{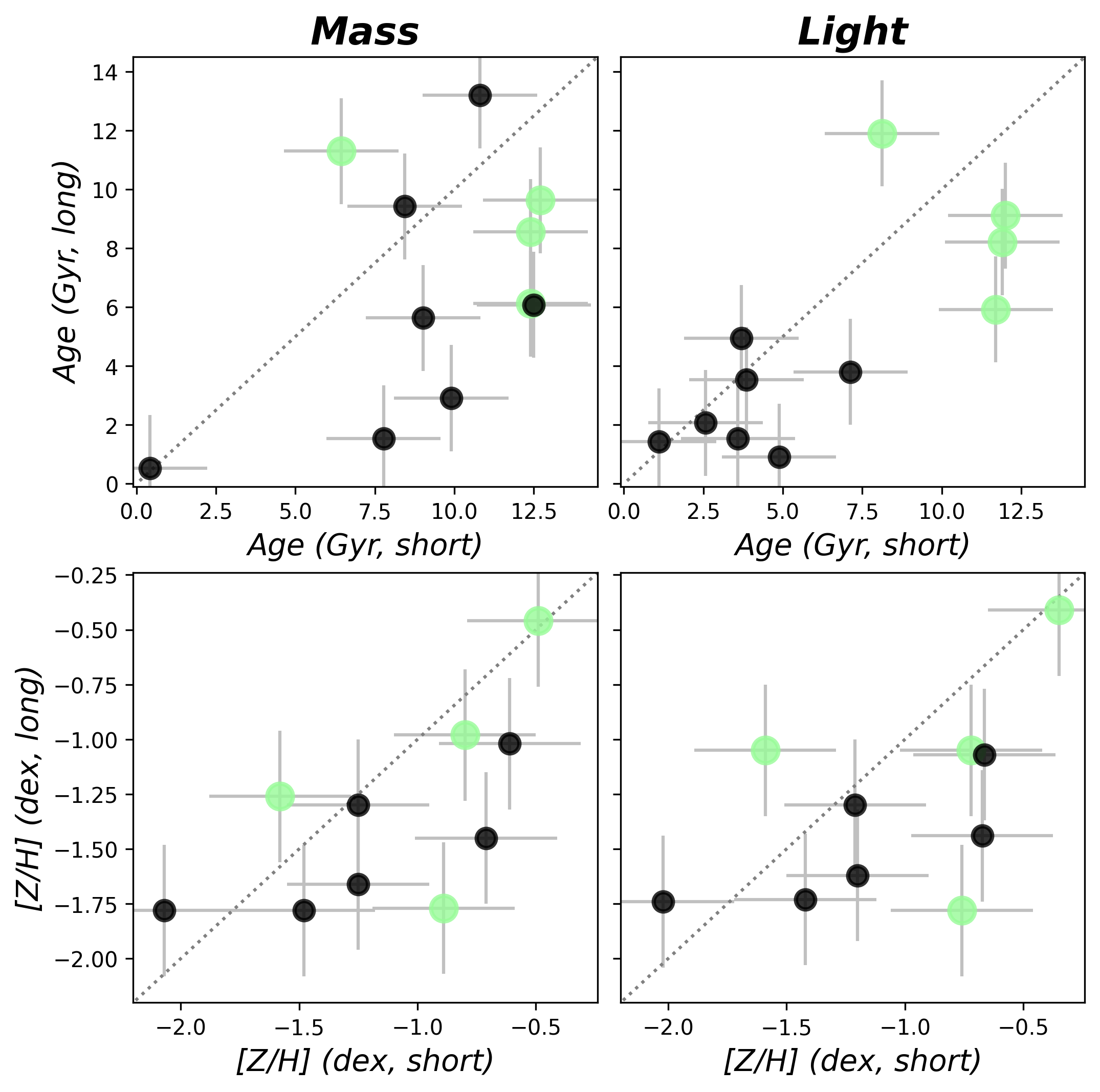}
\vspace{-0.4cm}
\caption{Comparison of using a long spectral range (as the KCWI BL grating) as opposed to a shorter one (BH) to test possible systematics in the derivation of the main stellar populations. Age (top) and metallicity (bottom) are shown, both for their mass-weighted (left panels) and light-weighted (right panels) values. Green points correspond to the UDGs observed with the BL configuration, which had their analysis repeated with the short baseline simulating the BH grating. Black symbols correspond to the GCs of \citet{Forbes2022}, where it was shown that missing the H$_{\beta}$ could bias the results towards younger ages. Overall, the biggest impact is seen only for the GCs that did not include the H$_{\beta}$.}
\label{fig:shortVSlong_fig}
\end{figure}

In \citet{Forbes2022} it was shown that for a sample of GCs observed with a short baseline that did not include H$_{\beta}$ (the main age indicator in the optical range), tend to bias the results towards younger ages. Although we have this line in all our spectra, we still want to test the impact on using a short (4800 to 5300\AA) as opposed to a long (3800 to 5500\AA) baseline from the different gratings. Therefore we repeat the fitting for PUDG-R27, PUDG-R84, DFX1 and DF07, limiting now the fit to the short spectral range. The results are shown in Figure \ref{fig:shortVSlong_fig}. We also include the GCs from \citet{Forbes2022}, where this was first studied for the KCWI gratings. We find that as long as H$_{\beta}$ is included, the recovered parameters are not much affected. 

\subsection{Comparison to other literature and SED fitting results}\label{section:ap_complit}
\begin{figure*}
\centering
\includegraphics[scale=0.49]{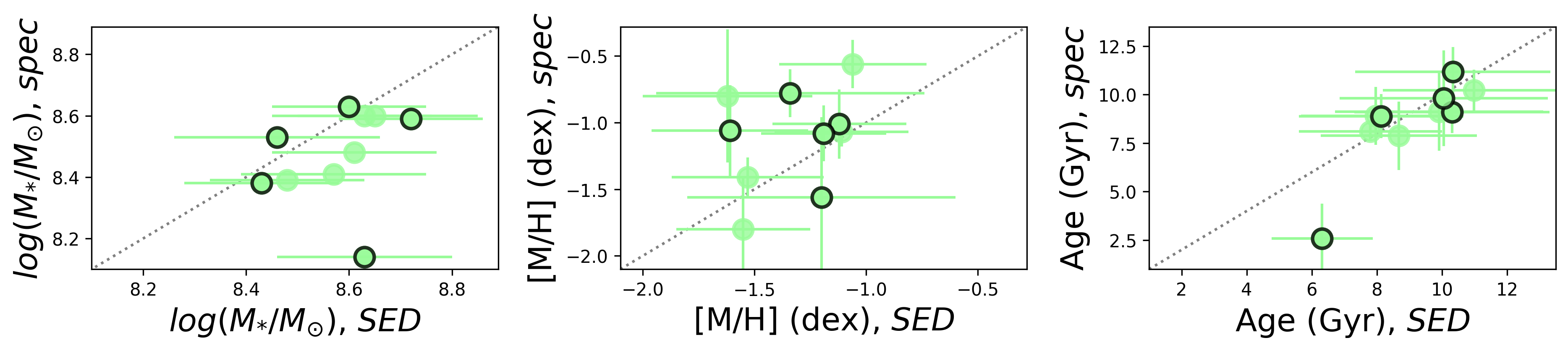}
\caption{Comparison of the obtained properties with those from the work of \citet{Buzzo2022} based on SED fitting. Green circles are the literature UDGs and the highlighted ones with black outline the new UDGs from this work. We present the derived stellar mass (left), metallicity (middle) and mass-weighted age (right). Overall, the stellar masses and ages are generally consistent. However, metallicities from SED fitting tend to be lower than those from spectroscopy, by roughly $\sim$0.25\,dex.}
\label{fig:sedVSspe_fig}
\end{figure*}

UDGs studies based on spectroscopy are still in their infancy given the challenges in obtaining large samples with high quality spectra. An alternative way to study some of the stellar populations is to obtain them from SED fitting, which tends to go deeper and is less time consuming. We next compare in Figure \ref{fig:sedVSspe_fig} the spectroscopic results of the UDGs in this work that have been previously analyzed using SED fitting by \citet{Buzzo2022}. In the latter it was shown that the solution that includes a small amount of dust tends to be more similar to the spectroscopic results, hence those are the values shown in the figure. Overall, both stellar masses and ages are compatible, but the stellar metallicities derived using the SED fitting provide systematically lower metallicities, by about 0.25\,dex (note however, that this is within the uncertainties). In fact, we find that the UDGs deviating the most in their metallicity tend to be the ones with very old stellar ages. 

\section{Plausible formation scenarios summary}\label{section:ap_summ}
\begin{table*}
\centering
\label{table:origins}                      
\begin{tabular}{c |c c c c c }   
\toprule      
\textbf{Galaxy}     &  phase-space & SFHs        &            Stellar Populations         &             RELATIONS             & GC richness   \\
                    &              &             &  Age / [M/H] / [Mg/Fe] & Age-[M/H] / [Mg/Fe]-[M/H] / MZR   &               \\
\midrule                                                                                     
\midrule                                                                                     
VCC~1287            &  VEI        & late+fast    &       O  / L / EE      &  dE-dSph / GC         /  within   &      R        \\
NGC~5846-UDG1       &  --         & late+fast    &       O  / L / EE      &  dE-dSph / GC         /  within   &      R        \\
UDG1137+16          &  --         & late+slow    &       Y  / L / SS      &  ?       / GC-dSph    /  below    &      --       \\
PUDG-R15            &  EI         & late+fast    &       EO / L / EE      &  GC-dE   / GC-dE      /  within   &      P        \\
PUDG-S74            &  mix        & early+slow   &       O  / H / E       &  dE      / dE         /  within   &      R        \\
PUDG-R84            &  mix        & late+fast    &       O  / L / E       &  dSph    / dSph-GC    /  below    &      R        \\
PUDG-R24            &  LI         & late+slow    &       Y  / L / US      &  dE      / dE         /  within   &      P        \\
PUDG-R27            &  VEI        & early+fast   &       EO / L / E       &  dE-GC   / dE         /  within   &      R        \\
Yagi358             &  VEI        & early+fast   &       O  / L / EE      &  dSph    / ?          /  below    &      R        \\
DFX1                &  mix        & late+fast    &       O  / L / EE      &  dE-dSp  / GC         /  within   &      R        \\
DF07                &  EI         & early+fast   &       EO / L / EE      &  GC-dE   / ?          /  within   &      R        \\
\midrule                                                                                                           
Yagi093             &  VEI        & early+slow   &       O  / H / E       &  dE      / dE         /  within   &      R        \\
Yagi098             &  VEI        & early+slow   &       I  / H / -       &  dE      / -          /  within   &      (R)      \\
Yagi275             &  LI         & late+slow    &       I  / H / US      &  dE      / dE         /  (above)  &      --       \\
Yagi276             &  VEI        & late+slow    &       I  / H / -       &  dE      / -          /  (above)  &      --       \\
Yagi392             &  VEI        & late+fast    &       I  / H / -       &  dE      / -          /  within   &      --       \\
Yagi418             &  VEI        & late+fast    &       O  / L / SS      &  dE-dSph / dE-dSph-GC /  within   &      (R)      \\
Yagi090             &  LI         & early+slow   &       I  / L / E       &  dE      / GC         /  within   &      (P)      \\
OGS1                &  VEI        & early+fast   &       O  / H / E       &  dE      / dE         /  within   &      (P)      \\
DF17                &  VEI        & --           &       O  / L / -       &  dE      / dE         /  within   &      R        \\
DF44                &  EI         & early+fast   &       EO / L / US      &  dSph-GC / dSph       /  below    &      R        \\
J130026.26+272735.2 &  VEI        & --           &       Y  / H / -       &  ?       / -          /  within   &      (P)      \\
J130038.63+272835.3 &  VEI        & --           &       Y  / H / -       &  dE      / -          /  within   &      (P)      \\
NGC~1052-DF2        &  --         & --           &       O  / L / SS      &  dE-dSph / dE-dSph    /  within   &      P        \\
DGSAT~I             &  --         & late+slow    &       O  / L / EE      &  dSph    / ?          /  below    &      (P)      \\
\bottomrule
\end{tabular}
\vspace{0.1cm}
\caption{\textbf{Summary of properties for each UDG}. \\
\textit{Column 2:} location information from phase-space (VEI - very early infall; EI - early infall; mix - mix regions; LI - late infall) \\
\textit{Column 3:} the type of SFH \\
\textit{Column 4:} main stellar populations properties: \\
- Stellar age (EO - extremely old, >10\,Gyr; O - old, 7.5 to 10\,Gyr; I - intermediate, 4.5 to 7.5\,Gyr; Y - young, <4.5\,Gyr); \\
- Metallicities (L - low, $\lesssim\,-$ 0.75\,dex; H, $\gtrsim\,-$0.75\,dex); \\
- $\alpha$-enhancement (EE - extremely enhanced, > 0.4\,dex; E - enhanced, 0.2 to 0.4\,dex; SS - scaled solar, 0.0 to 0.2 \,dex; US - under solar, <0.0\,dex)\\
\textit{Column 5:} type of object they resemble in the scaling relations and whether they follow or are located above or below the local MZR. \\
\textit{Column 6:} GC richness (in parenthesis for the visual estimated) \\
}
\end{table*}

We summarize in Table \ref{table:origins} all the properties discussed in this work. 
However, without further information on e.g. the dark matter content, total halo masses and proper GC counts, it is hard to point to a particular origin from the stellar populations alone. 

\bsp	
\label{lastpage}
\end{document}